\documentclass{article}
\usepackage[utf8]{inputenc}
\usepackage{amsmath}
\usepackage{amssymb}
\usepackage[margin=0.75in]{geometry}
\usepackage{hyperref}
\usepackage{cleveref}
\usepackage{indentfirst}
\usepackage{titlesec}
\usepackage{graphicx}
\usepackage[dvipsnames]{xcolor}
\usepackage{array}
\usepackage{multirow}
\usepackage{pifont}
\usepackage{hhline}
\allowdisplaybreaks

\newcommand{\del}{\partial}

\newcommand{\la}{\lambda}
\newcommand{\m}{\mu}
\newcommand{\n}{\nu}

\newcommand{\p}{\phi}
\newcommand{\vp}{\varphi}
\newcommand{\ve}{\varepsilon}
\newcommand{\td}{\text{d}}
\newcommand{\tm}{\tilde{M}}

\newcommand{\tg}{\tilde{g}}
\newcommand{\bp}{\bar{p}}
\newcommand{\ba}{\bar{\alpha}}
\newcommand{\bb}{\bar{\beta}}
\newcommand{\bi}{\bar{\iota}}
\newcommand{\bO}{\bar{\Omega}}
\newcommand{\bo}{\bar{\omega}}
\newcommand{\be}{\bar{e}}

\newcommand{\cmark}{\ding{51}}
\newcommand{\xmark}{\ding{55}}

\title {\bf Deformed black hole in Sagittarius A}
\date{}
\author{T. Anson $^{1}$, E. Babichev $^{1}$ and C. Charmousis $^{1}$ \\ \\
	$^{1}$ Universit\'e Paris-Saclay, CNRS/IN2P3, IJCLab, 91405 Orsay, France }

\begin{document}

	{\let\newpage\relax\maketitle}

	\begin{abstract}
We analyze the post-Newtonian orbit of stars around a deformed Kerr black hole. The deformation we consider is a class of disformal transformations of a nontrivial Kerr solution in scalar-tensor theory which are labeled via the disformal parameter $D$. We study different limits of the disformal parameter, and compare the trajectories of stars orbiting a black hole to the case of the Kerr spacetime in general relativity, up to 2PN order. Our findings show that for generic nonzero $D$, the no-hair theorem of general relativity is violated, in the sense that the black hole's quadrupole $Q$ is not determined by its mass $M$ and angular momentum $J$ through the relation $Q=-J^2/M$. Limiting values of $D$ provide examples of simple and exact noncircular metric solutions, whereas in a particular limit, where $1+D$ is small but finite, we obtain a leading correction to the Schwarzschild precession due to disformality. In this case, the disformal parameter is constrained using the recent measurement of the pericenter precession of the star S2 by the GRAVITY Collaboration. 
	\end{abstract}
	%%%%%%%%%%%%%%%%%%%%%%%%%%%%%%%%%%

	\section{Introduction}

In the center of the Milky Way lies Sagittarius~A (Sgr A), a complex radio source hidden at optical wavelengths by large clouds of optic dust.  Within Sgr~A, in the center of a spiral structure, lies the very bright compact astronomical radio and infrared source Sgr~A*. There is strong evidence that a supermassive black hole of mass $M_* \sim 4 \times 10^6 M_\odot$ inhabits the center of our Galaxy, and its location coincides with Sgr~A* (see Refs.~\cite{Ghez:2009yz,Abuter:2018drb,Abuter:2020dou} and references within).
The local supermassive black hole, for our purposes Sgr~A*, due to its proximity, of roughly 8 kpc, and strong gravity, provides a very promising avenue to test the general theory of relativity (GR) \cite{Johannsen:2015mdd}. Indeed, while GR has passed all observational  tests \cite{Will:2014kxa}, the gravitational field  around  black holes remained until very recently largely untested. In this regard, there have been considerable breakthroughs in recent years, whether it be by the detection of gravitational waves from a black hole merger \cite{Abbott:2016blz}, or the imaging of the supermassive black hole M87*'s shadow by the Event Horizon Telescope \cite{Akiyama:2019cqa}.  Additionally, the GRAVITY Collaboration \cite{refId0} has measured the redshift and pericenter precession of the star S2 orbiting Sgr~A*\cite{Abuter:2018drb,Abuter:2020dou}, and its results are compatible with the GR predictions. The GRAVITY measurements are the most recent and most precise addition to the observation of star orbits in the vicinity of Sgr~A*, which have been gathering radio and infrared observational data for more than two and a half decades~\cite{10.1117/12.317287,10.1117/12.321686}. In the future, the spin and eventually the quadrupole of Sgr~A*  could in principle be measured by the precise study of multiple star trajectories orbiting the black hole, especially those that have short periods of the order of a year \cite{Will:2007pp}. Pulsar timing measurements on the other hand would also provide a very promising alternate avenue to determine such parameters (see  e.g. Ref.~\cite{Psaltis:2015uza}) but for the moment their presence in the immediate vicinity of Sgr~A* remains elusive.  In GR, from the theoretical point of view a stationary and axially symmetric black hole is uniquely given by the Kerr metric \cite{Kerr:1963ud}, which depends on two parameters: the mass $M$ and angular momentum $J=a M$ of the black hole. According to the no-hair theorems all higher-order multipoles are then determined by these parameters. Notably, the quadrupole moment $Q$ satisfies $Q=-J^2/M$, so an independent measurement of $J$ and $Q$ using star orbits or pulsar timing measurements would provide an excellent test of the black hole nature of Sgr~A*.

In this rapidly evolving observational context, it is important to study rotating black hole metrics alternative to Kerr. This would provide viable rulers by which departures from GR can be quantified and eventually a starting point to charter new phenomena. Given the difficulty of finding exact solutions describing rotating black holes, several numerical solutions have been constructed, using powerful numerical techniques (see for example Refs.~\cite{Herdeiro:2014goa, Vincent:2016sjq, Vincent:2015xta, VanAelst:2019kku} and references within). On the other hand the usual analytical approach is, starting from some interesting geometric property (integrability of geodesics, proximity to Kerr, etc.), to construct metric deformations of the Kerr spacetime from a theory-agnostic point of view;  see Ref.~\cite{Johannsen:2015mdd} for examples of such deformations. Although this approach is quite powerful, such \emph{ad hoc} deformations, are not solutions of some theory and may lead to singular or noncausal spacetimes \cite{Johannsen:2013rqa}.

In the framework of degenerate higher-order scalar-tensor (DHOST) theories of gravity, \cite{Langlois:2015cwa,Crisostomi:2016czh},  regular deformations of the Kerr metric were constructed recently in Ref.~\cite{Anson:2020trg} (see also Ref.~\cite{BenAchour:2020fgy}). This was technically achieved using a disformal transformation of a stealth Kerr solution of DHOST theory~\cite{Charmousis:2019vnf}. In the absence of a better name, we will often refer to these disformal transformations of Kerr as {\it{disformed Kerr}} spacetimes. Their departure from Kerr is parametrized by a constant $D$. Each constant $D$ corresponds to a {\it different} disformed metric and also labels a {\it different} class of DHOST theories that admit the disformed metric as a solution. In this sense, the disformal parameter $D$ does not correspond to extra hair (because changing $D$ changes the theory), but rather it labels a new class of black hole solutions in modified gravity theories. Without being agnostic on the theory, we can also follow a phenomenological point of view, and consider $D$ as an extra parameter, and even study some of its  limits. 
Interestingly, for the disformed metric, the scalar field acts as a global time function, ensuring that no closed timelike curves appear in the region outside the event horizon ~\cite{Anson:2020trg}.
Furthermore, these solutions were shown to have many interesting properties which differentiate them from the Kerr metric: the event horizon no longer lies at constant radius $r$ and is not a Killing horizon; the limiting surface for stationary observers is generically distinct from the outer event horizon; the metric is noncircular, meaning that it cannot be written in a form that exhibits the reflection symmetry $(t,\vp)\to(-t,-\vp)$; see Ref.~\cite{Anson:2020trg} for details. The latter property is usually assumed for axisymmetric spacetimes in the literature as it is a property of vacuum GR solutions. Therefore, already, the disformed Kerr metrics are interesting counterexamples and their study could lead to a better understanding of the implications of noncircularity. Even though the geodesic equation is not separable in this case, one can integrate it numerically, and the shadows of disformed Kerr black holes were studied in Ref.~\cite{Long:2020wqj}.

In this work, we will assume that the gravitational field of Sgr~A* is described by the disformed Kerr metric, and study the bound post-Newtonian (PN) orbits of stars in this spacetime. In this case one can expand the metric asymptotically in powers of $M/r$, where $r$ is the distance of the star to the black hole. We will consider different limiting cases for the disformal parameter $D$. When $D=0$, we have a Kerr black hole and the secular evolution of orbital parameters is well known (see for instance Ref.~\cite{Will:2016pgm}). The precession of the 
star's pericenter is a 1PN effect, and already exists for the Schwarzschild spacetime. When the black hole is spinning, we obtain additional effects at higher PN orders. The dragging of inertial frames (or Lense-Thirring effect) appears at 1.5PN order, while quadrupole contributions appear at 2PN order. We will examine how the predictions of the disformal Kerr metric differ from those of the Kerr spacetime for different values of $D$, and up to 2PN order. We will see that generically the no-hair theorem is violated in the disformal case, which makes these metrics interesting counterexamples to the Kerr spacetime. Additionally, we will see that in one case, namely $1+D\sim M/r$, the Schwarzschild precession is modified by disformality, and hence it is already possible to constrain the disformal parameter using current observations in this case.

The paper is structured as follows. In Sec.~\ref{sec:disformal}, we discuss the asymptotic properties of the disformed Kerr metric~\cite{Anson:2020trg}. In Sec.~\ref{sec:method}, we present the osculating orbit method used in this work to calculate the secular evolution of orbital parameters, following Ref.~\cite{Will:2016pgm}. In Sec.~\ref{sec:orbit_pert}, we analyze the secular evolution of orbital parameters for various cases of the disformal parameter $D$. We start with the generic case, where $D$ is neither too large nor too close to $-1$, before considering the limits $D\to\infty$, $D\to -1$, and the case where $D+1$ is small but finite separately. Finally, we discuss observational constraints and summarize our results in Sec.~\ref{sec:conclusion}.
	
	%%%%%%%%%%%%%%%%%%%%%%%%%%%%%%%%%%%%%%%%%%%%%%%%%%%%%%%%
	\section{The disformed Kerr metric}
	\label{sec:disformal}
	%%%%%%%%%%%%%%%%%%%%%%%%%%%%%%%%%%%%%%%%%%%%%%%%%%%%%%%%%%%%%%
	The starting point of our study is the following deformation (or more exactly disformal transformation) of the Kerr metric~\cite{Anson:2020trg}:
	\begin{equation}
	\begin{split}
	\label{metric}
	\tilde{g}_{\mu\nu}\td x^\mu \td x^\nu &= -\left(1-\frac{2\tilde{M}r}{\rho^2}\right)\td t^2 -\frac{4\sqrt{1+D}\tilde{M}a r\sin^2\theta}{\rho^2}\td t\td\vp + \frac{\sin^2\theta}{\rho^2}\left[\left(r^2+a^2\right)^2-a^2\Delta
	\sin^2\theta\right]\td\vp^2\\
	&+  \frac{\rho^2 \Delta - 2 \tilde{M}(1+D) r D (a^2+r^2)}{\Delta^2}\td r^2 - 2D \frac{\sqrt{2\tilde{M}r(a^2+r^2)}}{\Delta}\td t\td r + \rho^2 \td \theta^2\; ,
	\end{split}
	\end{equation}
	where $\tilde{M}$ is the mass of the black hole, $a$ is related to the angular momentum\footnote{The value of the angular momentum per unit mass as seen by an asymptotic observer is actually $\tilde{a} = a\sqrt{1+D}$, as explained below.}, and we have introduced the shorthand notations, 
	\begin{equation*}
	\Delta = r^2+a^2 - 2 Mr\; , \quad \rho^2 = r^2 + a^2 \cos^2\theta\; ,
	\end{equation*}
where we note that $M=(1+D)\tm$.
	The deformation of the Kerr metric is encoded in the constant parameter $D$ in Eq.~(\ref{metric}).
	For each $D$,  the metric~(\ref{metric}) with the associated scalar field is an exact solution of a given subclass of DHOST theory (see
	Appendix~\ref{app:disformed} for details).  Given the phenomenological scope of this paper, we will treat $D$ as a deformation parameter, assuming in particular that $1+D\geq 0$ for the disformed metric to be real. In the following, we assume that matter fields are minimally coupled to the disformed metric. Being in the physical frame, the scalar field does not directly couple to matter and therefore does not create the fifth force.
	
	One can easily check that for $D=0$ the metric~(\ref{metric}) reduces to the Kerr metric in Boyer-Lindquist (BL) coordinates.
	A nontrivial property of the disformed metric is the presence of an off-diagonal term $\tilde{g}_{tr}$, which cannot in general be eliminated by a coordinate change without introducing other off-diagonal elements; see the discussion in Ref.~\cite{Anson:2020trg}. 
	This can be linked to the fact that for $D\neq 0$ the metric~(\ref{metric}) is noncircular, in contrast to the Kerr metric. 
	An exception is the static case $a=0$, for which it can be shown by an appropriate coordinate transformation that the metric is simply that of a Schwarzschild black hole of mass $\tm$ \cite{Babichev:2017lmw,Babichev:2018uiw}.

	While the strong gravity regime is very different from that of the Kerr metric~\cite{Anson:2020trg}, the asymptotic form of Eq.~(\ref{metric}) is similar to the GR spacetime for generic values of $D$. To see this, we first perform the following coordinate change:
	\begin{equation}
	\label{time_redef}
	\td t \to \td t - D\frac{\sqrt{2\tm r(a^2 + r^2)}}{\Delta(1-\frac{2\tm}{r})}\td r\; .
	\end{equation}
	When $a=0$, this is the coordinate change that brings the metric to the familiar Schwarzschild coordinates. In the general case, this transformation makes the $\tilde{g}_{tr}$ term smaller in orders of $\tilde{M}/r$, so that crucially the leading asymptotic terms are the same as those of Kerr in BL coordinates.
	Indeed, after this coordinate change (\ref{time_redef}), the asymptotic form of the disformed metric is
	\begin{equation}
	\label{asymptotic} 
	\begin{split}
	\td \tilde{s}^2& = -\left[1-\frac{2\tilde{M}}{r} + \mathcal{O}\left(\frac{1}{r^3}\right)\right]\td t^2 -
	\left[\frac{4 \tilde{a} \tilde{M}}{r^3}+ \mathcal{O}\left(\frac{1}{r^5}\right)\right]\left[x \td y - y\td x\right]\td t+ \left[\delta_{ij} + \mathcal{O}\left(\frac{1}{r}\right)c_{ij} \right]\td x^i\td x^j \\
	&+\frac{D}{1+D}\left[\mathcal{O}\left(\frac{\tilde{a}^2\tilde{M}}{r^3}\right)\td
	t^2 +
	\mathcal{O}\left(\frac{\tilde{a}^2\tilde{M}^{3/2}}{r^{7/2}}\right)b_i\td
	t\td x^i +
	\mathcal{O}\left(\frac{\tilde{a}^2}{r^{2}}\right)d_{ij}\td
	x^i\td x^j\right]\; ,
	\end{split}
	\end{equation}
	where $\tilde{a} = a \sqrt{1+D}$, and $\{b_i,c_{ij},d_{ij}\}\sim\mathcal{O}(1)$. 
	It can be seen from the above expression that the values of the observed mass and angular momentum are given by $\tilde{M}$ and $\tilde{a}$ correspondingly. In other words, the deformation parameter $D$ affects the mass and angular momentum of the black hole. 
	Apart from this important fact, the disformal corrections only affect subleading terms of the asymptotic Kerr spacetime. 
	However, when $D+1\to 0$, the higher-order corrections become important, as we will see in the following. 
	Though it is not obvious from Eq.~(\ref{asymptotic}), we will also see that the limit $D\to \infty$ gives rise to a notable difference from the Kerr spacetime asymptotically.
	
	Our goal is to study the post-Newtonian motion of stars in the vicinity of a disformed Kerr spacetime. 
	In the Newtonian limit the trajectory of a star forms an ellipse with the black hole located at one of its foci. 
	In order to describe the post-Newtonian motion, we introduce the following dimensionless parameters:
	\begin{equation}
	\varepsilon=\frac{\tilde{M}}{A}\; ,\qquad \chi = \frac{\tilde{a}}{\tilde{M}}\; ,
	\end{equation}
	where $A$ is the semimajor axis of the ellipse, and the dimensionless spin $\chi$ is assumed to be of $\mathcal{O}(1)$.
	We consider the case where the star is far away from the black hole, and write the metric up to 2PN order, meaning that we keep terms up to $\mathcal{O}(\varepsilon^3)$. As stated above, we first perform the coordinate change (\ref{time_redef}) so that the asymptotic form of the metric is closer to Kerr in BL coordinates. After this redefinition, the line element up to 2PN order reads
	\begin{equation}
	\label{metric_as}
	\begin{split}
	\td\tilde{s}^2_\text{2PN}&=-\left(1-\frac{2 \tm}{r}+\frac{2 \tm^3\chi^2\cos^2\theta }{(1+D)r^3}\right)\td t^2 + \left(1+\frac{2\tm}{r}+\frac{4\tm^2}{r^2}-\frac{\tm^2\chi^2\sin^2\theta}{(1+D)r^2}\right)\td r^2\\
	&+r^2\left(1+\frac{\tm^2\chi^2\cos^2\theta}{(1+D)r^2}\right)\td\theta^2+r^2\sin^2\theta\left(1+\frac{\tm^2\chi^2}{(1+D)r^2}\right)\td\vp^2 -\frac{4\tm^2\chi\sin^2\theta}{r}\td t\td\vp\; .
	\end{split}
	\end{equation}
	Note that we keep terms up to $\mathcal{O}(\varepsilon^3)$ in the $\tg_{tt}$ component, and lower-order terms in $\varepsilon$ in other components because the motion of stars is assumed to be nonrelativistic. Indeed, for the motion of a nonrelativistic star, the spatial variation is suppressed with respect to the time variation along the trajectory by the 3-velocity  $v\sim \sqrt{\varepsilon}$, i.e.  $\td x^i\sim \sqrt{\varepsilon} \td t$, and therefore one needs to keep lower-order terms in the spatial components of the metric.
	At this PN order, the metric is circular (meaning it is unchanged under the reflection $(t,\vp)\to(-t,-\vp)$), and the expansion is very close to the Kerr metric. 
	One can also check that for Eq.~(\ref{metric_as}) the Ricci tensor is nonzero only at $\varepsilon^3$ order, i.e. $R_{\mu\nu}\sim \mathcal{O}(\varepsilon^3)$ (in these coordinates).
	This can be seen by evaluating the Ricci tensor for the full metric~(\ref{metric}) (see also Ref.~\cite{Anson:2020trg} where the curvature invariants were computed for Eq.~(\ref{metric})).
	Thus one can say that the metric~(\ref{metric_as}) is Ricci-flat up to the order $\varepsilon^2$.

Once we have read off the mass and spin of the black hole from the $\tg_{tt}$ and $\tg_{t\vp}$ terms, the disformal factor $D$ only enters the terms proportional to $\chi^2$ in Eq.~(\ref{metric_as}), which correspond to quadrupole terms\footnote{It is worth noting that these terms correspond to the leading-order contributions of the Newtonian quadrupole moment, even though we will refer to them as 2PN in the context of a large-$r$ expansion.}. 
	In other words, the disformal metric is equivalent to the Kerr metric up to 1.5PN order for generic $D$.
	To better understand the form of deviations at higher PN orders, it is  instructive to compare the metric~(\ref{metric_as}) to a non-Kerr metric at that order.
	A particularly interesting example is the Butterworth-Ipser (BI) metric~\cite{BI75,BI76}, which was constructed to model a rapidly rotating star. The BI metric is usually written in quasi-isotropic coordinates (see Appendix \ref{app:secular}). 
	Therefore in order to see a relation of the asymptotic expansions of the disformed Kerr and the BI metrics, we write the 2PN-order expansion of the BI metric in BL-like coordinates. The coordinate change is given in Appendix~\ref{app_others} (see Eq.~(\ref{QItoBL})), which results in
		\begin{equation}
		\label{BI_BL}
		\begin{split}
		\td s^2_\text{BI}&=-\left(1-\frac{2 \tm}{r}+\frac{2 \tm^3}{r^3}\left[q + 6 a_0 + \frac{\chi^2}{2(1+D)}-3(4 a_0 + q)\cos^2\theta\right]\right)\td t^2 -\frac{4\tm^2\chi\sin^2\theta}{r}\td t\td\vp\\
		& + \left(1+\frac{2\tm}{r}+\frac{4\tm^2}{r^2}+\frac{\tm^2}{r^2}\left[6a_0\cos2\theta - \frac{\chi^2}{2(1+D)}\right]\right)\td r^2\\
		&+r^2\left(1+\frac{\tm^2
		}{r^2}\left[6a_0 \cos 2\theta + \frac{\chi^2}{2(1+D)}\right]\right)\td\theta^2+r^2\sin^2\theta\left(1+\frac{\tm^2
	}{r^2}\left[6 a_0 + \frac{\chi^2}{2(1+D)}\right]\right)\td\vp^2 \; ,
		\end{split}
		\end{equation}
where $a_0$ and $q$ are quadrupole parameters and we adopt here the notations of Ref.~\cite{Friedman:2013xza} (but we use $a_0$ instead of their $a$ to avoid confusion with the Kerr spin parameter). In fact with these conventions, the parameter $q$  corresponds to the coordinate-invariant quadrupole moment, as pointed out in Ref.~\cite{Pappas:2012ns} (see also Ref.~\cite{Paschalidis:2016vmz}).
One can see that the disformed Kerr and the BI metrics can indeed be matched at 2PN order. 
A direct comparison of the above line element to the metric~(\ref{metric_as}) gives the following identification of the parameters for the disformed metric in the case of generic $D$:
\begin{equation}
\label{BI_param}
a_0^{(D)} = \frac{\chi^2}{12(1+D)}\; ,\qquad q^{(D)} = -\frac{\chi^2}{1+D}\; ,
\end{equation}

It may seem surprising that such a matching exists, taking into account the completely different nature of the disformed Kerr and BI metrics, and given that there are only two free parameters at hand. It should be noted however, that at higher PN orders, where the noncircularity of the disformed metric (re)appears,  such a matching cannot be done as the BI metric is circular (see for example Refs.~\cite{Gourgoulhon:2010ju, Friedman:2013xza}). \footnote{Noncircularity appears in GR neutron stars when considering fluids with convective meridional currents (see for example Refs.~\cite{PhysRevD.48.2635, Birkl:2010hc}).} Furthermore, for some limiting cases of $D$ that we will consider below, the matching does not exist, i.e. the disformed metric cannot be written in the BI form, even at 2PN order.

In the following, we will apply the theory of orbital perturbations to calculate the secular variation of orbital parameters. 
	As we will see, the secular evolution of orbital parameters does not change up to 1.5PN order for generic $D$, as can be guessed from the form of the expansion~(\ref{metric_as}).
	However, for small enough $(1+D)$ modifications of the time variation of orbital parameters happen at lower orders. 
	In this case, the experimental bounds coming from the GRAVITY Collaboration can be used to constrain the disformal parameter.
	In the following we will study various parameter ranges of $D$ that may lead to testable effects.
	
	%%%%%%%%%%%%%%%%%%%%%%%%%%%%%%%%%%%%%%%%%%%%%%%%%%%%%%%%%%%%%%%
	\section{Osculating orbit method}
	%%%%%%%%%%%%%%%%%%%%%%%%%%%%%%%%%%%%%%%%%%%%%%%%%%%%%%%%%%%%%%%
	\label{sec:method}
	\begin{figure}[t]
		\centering
		\includegraphics[scale=0.35]{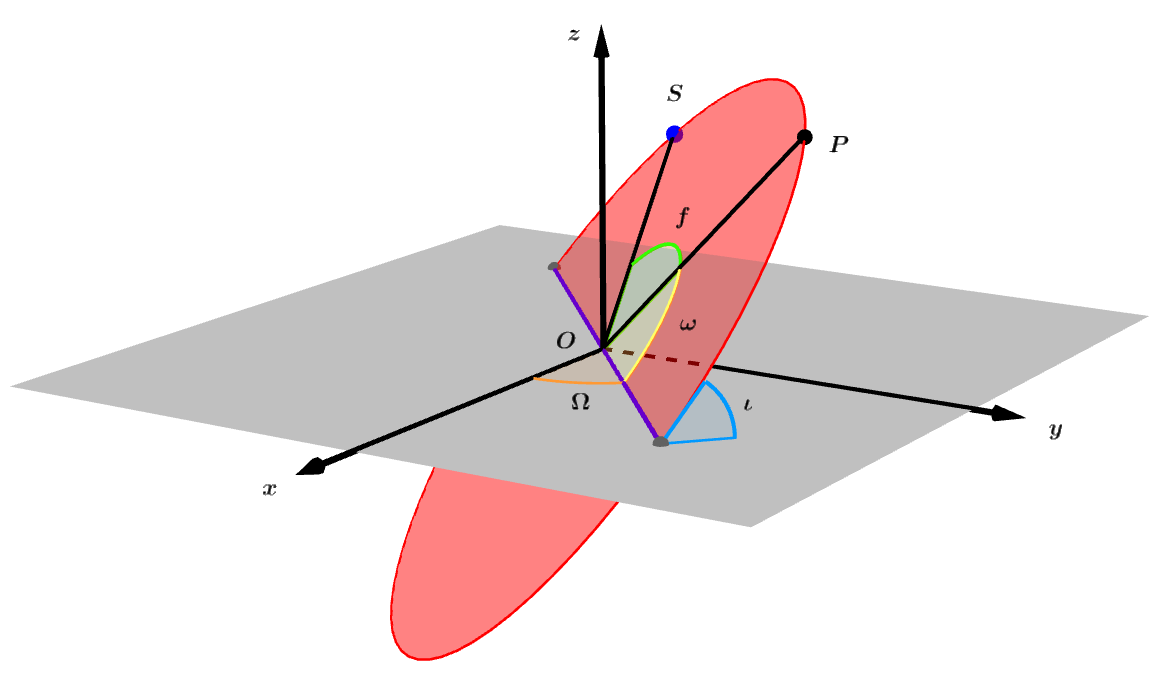}
		\caption{Kepler orbit of a star $S$ around the black hole located at $O$. The purple line is called the line of nodes, and it is defined by the intersecting points of the star's trajectory with the $(Oxy)$ reference plane. The nodal angle $\Omega$ gives the position of this line with respect to the $(Ox)$ axis. Starting from the line of nodes, the pericenter $P$ of the trajectory is given by the pericenter angle $\omega$, while $\iota$ represents the inclination angle of the ellipse with respect to the  $(Oxy)$ plane. Finally, the true anomaly, $f$, gives the position of the star S with respect to the pericenter $P$.}
		\label{angle_def}
	\end{figure}
	We will use the standard osculating orbit method  to compute the secular variation of orbital elements (see for instance Ref.~\cite{Soffel:2019aoq}).
	In general, the three-dimensional acceleration of a test body can be written in the following form:
	\begin{equation}
	\label{acceleration}
	\textbf a = -\frac{\tm}{r^3}\textbf{x} + \textbf F\; ,
	\end{equation}
	where the first term on the right-hand side corresponds to the Newtonian acceleration, $\textbf F$ is the perturbation of the Newtonian acceleration and $\textbf{x}$ is the position vector in space, so that $r=|\textbf{x}|$. 
	We need to calculate the projections of the acceleration~(\ref{acceleration}) along the orthogonal directions $\textbf{x}$, $\textbf{h}= \textbf{x}\times\textbf{v}$, and $\textbf{h}\times\textbf{x}$, where $\textbf{v}=\td\textbf{x}/\td t$ is the 3-velocity of the star. These projections are given by
	\begin{equation*}
	\mathcal{S} =\frac{1}{r} \textbf{x}\cdot\textbf{F}\; , \qquad \mathcal{T} = \frac{1}{h r}(\textbf{h}\times\textbf{x})\cdot\textbf{F}\; ,\qquad \mathcal{W} = \frac{1}{h} \textbf{h}\cdot\textbf{F}\; ,
	\end{equation*}
	where $h=|\textbf{h}|$.
	The expressions for the components of $\textbf{x}$ in Cartesian coordinates (see Fig.~\ref{angle_def}) with respect to the orbital elements read
	\begin{equation*}
	\begin{split}
	x &= r\left[\cos\Omega\cos u - \sin\Omega\cos \iota \sin u\right]\; ,\\
	y &=r\left[\sin\Omega\cos u + \cos\Omega\cos \iota \sin u\right]\; ,\\
	z&= r \sin \iota \sin u\; ,
	\end{split}
	\end{equation*}
	where $ u=\omega+f$.
	To obtain the components of $\textbf{v}$ one differentiates the above expression assuming that all the angles are constant except for $u$. 
	In the following, we use the standard relations,
	\begin{equation}
	r = \frac{p}{1+e\cos f}\; ,\qquad\frac{dr}{dt} = \frac{e h}{p}\sin f\; ,\qquad h = \sqrt{\tm p}\; , \qquad p = A\left(1-e^2\right)\; ,
	\end{equation}
	where $A$ and $e$ are respectively the semimajor axis and eccentricity of the ellipse.
	In order for the limit $e\to0$ to be well defined, the alternative orbit parameters $\alpha = e \cos\omega$ and $\beta= e \sin\omega$ are introduced.
	With the above definitions, we can write the evolution equations for all the orbital parameters (see for instance Ref.~\cite{Will:2016pgm}, though notice a typo in the expressions for $\td\alpha/\td t$ and $\td\beta/\td t$):
	\begin{equation}
	\label{time_evo}
	\begin{split}
	\frac{\td p}{\td t} &=2r\sqrt{\frac{p}{\tm}}\mathcal{T}\; ,\\
	\frac{\td \alpha}{\td t} &=\sqrt{\frac{p}{\tm}}\left[\mathcal{S}\sin u+\left(\frac{\alpha r}{p}+\left(1+\frac{r}{p}\right)\cos u\right)\mathcal{T}+\frac{\beta r \mathcal{W}}{p}\cot \iota\sin u\right]\; ,\\
	\frac{\td \beta}{\td t} &=\sqrt{\frac{p}{\tm}}\left[-\mathcal{S}\cos u+\left(\frac{\beta r}{p}+\left(1+\frac{r}{p}\right)\sin u\right)\mathcal{T}-\frac{\alpha r \mathcal{W}}{p}\cot \iota\sin u\right]\; ,\\
	\frac{\td \iota}{\td t} &=\frac{r\mathcal{W}\cos u}{\sqrt{\tm p}}\; ,\\
	\frac{\td \Omega}{\td t} &=\frac{r\mathcal{W}\sin u}{\sqrt{\tm p}\sin \iota}\; ,\\
	\frac{\td u}{\td t} &= \frac{h}{r^2}-\cos \iota\frac{\td\Omega}{\td t}\; .
	\end{split}
	\end{equation}
	Using the evolution equations~(\ref{time_evo}), we follow the analysis of Ref.~\cite{Will:2016pgm} to obtain the secular variation of orbital elements. 
	First, we use the last equation of Eq.~(\ref{time_evo}) to trade $dt$ for $du$ in all other equations and write them in the form,
	\begin{equation*}
	Q_k = \frac{\td X_k}{\td u}\; ,
	\end{equation*}
	where the $X_k$ stands for orbital parameters $p$, $\alpha$, $\beta$, $\iota$ and $\Omega$. We now perform a two-timescale analysis  \cite{Hinderer:2008dm,bender2013advanced,Lincoln:1990ji,Mora:2003wt,Will:2016pgm} by introducing a second variable $\Theta=\epsilon u$, where $\epsilon$ is a bookkeeping parameter that is useful to keep track of small terms. Since $\Theta$ varies on longer timescales, in the following we treat $u$ and $\Theta$ as independent variables. 
	This approach allows us to make an average over a period using the variable $u$, while keeping $\Theta$ as a slowly-varying (almost constant) variable.
	We define the average holding $\Theta$ fixed as
	\begin{equation}
	\label{average}
	\langle C\rangle =\frac{1}{2\pi}\int_{0}^{2\pi}C(\Theta,u)\,\td u\; ,
	\end{equation} 
	and each orbital parameter is decomposed as
	\begin{gather}
	X_k(\Theta,u) = \bar{X}_k(\Theta)+\epsilon Z_k(\bar{X}_l,u)\; ,\\
	\bar{X}_k(\Theta)=\langle X_k(\Theta,u)\rangle\; ,\qquad\langle Z_k(\bar{X}_k(\Theta),u)\rangle=0\; .
	\end{gather}
	This analysis is not necessary to obtain leading-order terms in the variation of orbital parameters, but the periodic contributions which appear in the $Z_k$ must be taken into account if one calculates higher-order terms in $\epsilon$.
	In order to obtain secular variations of the orbital parameters, one can derive the following formula (see Ref.~\cite{Will:2016pgm} for details):
	\begin{equation}
	\frac{\td \bar{X}_k}{\td u}=\epsilon\langle Q^0_k\rangle + \epsilon^2\left[\langle Q_{k,l}^0\int_{0}^{u}Q_l^0\td u'\rangle + \langle Q^0_{k,l}\rangle\langle u Q^0_{l}\rangle-\langle \left(u + \pi\right)Q^0_{k,l}\rangle\langle  Q^0_{l}\rangle\right]+\mathcal{O}\left(\epsilon^3\right)\; ,
	\end{equation}
	where $Q^0_k= Q_k(\bar{X}_l,u)$. In the following section we will make use of this approach to find the secular orbital shifts for different ranges of the parameters of the disformed metric~(\ref{metric}).

	%%%%%%%%%%%%%%%%%%%%%%%%%%%%%%%%%%%%%%%%%%%%%%%%%%%%%%%%%%%%%%%%%%
	\section{Orbital perturbations for the disformed Kerr metric}
	%%%%%%%%%%%%%%%%%%%%%%%%%%%%%%%%%%%%%%%%%%%%%%%%%%%%%%%%%%%%%%%%%%
	\label{sec:orbit_pert}
	In this section we use the technique described above to calculate the secular shifts of orbital parameters for different cases involving the disformal  parameter $D$. 
	We will start with a generic $D$, and then consider limiting cases that provide interesting phenomenology.
	As is common in the literature, we coordinate transform to Kerr harmonic coordinates, i.e. coordinates verifying $\square x_\text{H}^\mu=0$, 
	where the $\Box$ operator corresponds to the Kerr metric with Kerr parameters $\tilde{M}$ and $\tilde{a}$.
	It should be made clear that
	one has $\tilde{\square}x_\text{H}^\mu\neq 0$, which means that
	these coordinates are {\it not } harmonic for the disformed metric.
	The idea, however, is to use the same coordinates that one uses when assuming the Kerr black hole and GR, in order to better gauge the differences arising from the disformed spacetime\footnote{We thank Gilles Esposito-Far\`ese for enlightening discussions concerning this issue.}. This is also the reason why we work with the black hole parameters $\{\tilde{a},\tm\}$ determined from the asymptotic expansion. 
	Therefore, this choice of coordinates makes it easier to link our results to observations.
	
	Below we first consider the case of generic  $D$ in Sec.~\ref{case_generic}, 
	and then study different limits of the parameter $D$.
	In particular, we investigate the large-$D$ limit in Sec.~\ref{case_infty}, for which  the metric~(\ref{metric}) and its asymptotic expansion~(\ref{metric_as}) simplify considerably, since a number of terms drop out.  The exact limit $D\to -1$ studied in Sec.~\ref{case_-1} provides another simple and interesting example. Finally, we study the limit of small but finite $(1+D)$, for which deviations from the Kerr geometry are \emph{enhanced} and, consequently, the corrections to orbital shifts become larger. 
		We consider two different regimes separately: $(1+D)\sim \varepsilon$ in Sec.~\ref{case_1+D_small}, and $(1+D)\sim \sqrt{\varepsilon}$ in Sec.~\ref{case_1+D_small2}. 
	
	%%%%%%%%%%%%%%%%%%%%%%%%%%%%%%%%%%%%%%%%%%%%%%%%%%%%%%%%%%%%%%%%%
	\subsection{Disformal Kerr: Generic case}
	\label{case_generic}
	We start with the generic case, where $D$ is arbitrary, but not too large or too close to $-1$. 
	In terms of the BL coordinates $\{t, r,\theta,\vp\}$, the harmonic coordinates $x^\mu_\text{H}$ can be written as \cite{Will:2016pgm,Harmonic}
	\begin{equation}
	\label{Kerr_harm}
	\begin{split}
	t_\text{H}&=t\; ,\\
	x_\text{H}&= \sqrt{\tilde{R}^2+a^2}\cos\Psi\sin\theta\; ,\\
	y_\text{H}&= \sqrt{\tilde{R}^2+a^2}\sin\Psi\sin\theta\; ,\\
	z_\text{H}&= \tilde{R}\cos\theta\; ,
	\end{split}
	\end{equation}
	where $\tilde{R}$ and $\Psi$ are defined as
	\begin{equation}
	\begin{split}
	\tilde{R}&=r-\tm\; ,\\
	\Psi& = \vp + \tan^{-1}\left(\frac{\tilde{a}}{r-\tm}\right) + \int \frac{\tilde{a}}{\Delta}\td r\; .
	\end{split}
	\end{equation}
	We now invert these relations up to $\mathcal{O}(\varepsilon^3)$, and replace the BL-like coordinates of Eq.~(\ref{metric_as}) by harmonic coordinates (we drop the index ``H" in the following). 
	The result is
	\begin{equation}
	\label{harm_no_lim}
	\begin{split}
	\tilde{g}_{00}&= -1+\frac{2\tm}{r}-\frac{2\tm^2}{r^2}+\frac{2\tm^3}{r^3} +\frac{\tm^3\chi^2}{r^3}\left[1-\frac{3+D}{1+D}\left(\textbf{n}\cdot\textbf{s}\right)^2\right]+\mathcal{O}\left(\varepsilon^4\right)\; ,\\
	\tg_{0j}&=\frac{2\tm^2\chi}{r^2}\left(\textbf{n}\times\textbf{s}\right)_j+ \mathcal{O}\left(\varepsilon^3\right)\; ,\\
	\tg_{ij}&=\left[1+\frac{2\tm}{r}+\frac{\tm^2}{r^2}\left(1-\frac{D\chi^2}{1+D}\right)\right]\delta_{ij}+\frac{\tm^2 }{r^2}n_i n_j +\frac{D\tm^2\chi^2}{(1+D)r^2}\left[2 n_i n_j+s_i s_j-2s_{(i}n_{j)}(\textbf{n}\cdot\textbf{s})\right]+\mathcal{O}\left(\varepsilon^{5/2}\right)\; ,
	\end{split}
	\end{equation}
	where $\textbf{n}=\textbf{x}/r$, $\textbf{s}=\textbf{J}/J=\textbf{e}_z$, and $r=r(x,y,z)$ is now the radial coordinate in the old metric expressed in harmonic coordinates.
	When $D=0$, the expressions~(\ref{harm_no_lim}) reduce to the Kerr metric components in harmonic coordinates (see Ref.~\cite{Will:2016pgm}). 
	
	We now apply the method described in Sec.~\ref{sec:method} to the metric (\ref{harm_no_lim}). 
	For generic values of $D$, the secular variation of orbit elements up to 2PN order is given by
	\begin{equation}
	\label{sec_general}
	\begin{split}
	\frac{\td \bp}{\td u}& = 0\; ,\\
	\frac{\td\ba}{\td u}& = -\frac{3\tm \bb}{\bp}+6\chi\bb \cos\bi\left(\frac{\tm}{\bp}\right)^{3/2}+\frac{3\tm^2\bb}{4 \bp^2}\left(10-\ba^2-\bb^2\right)-\frac{3\tm^2\bb\chi^2}{4\bp^2(1+D)}\left(5\cos^2\bi-1\right)\; ,\\
	\frac{\td\bb}{\td u}& = \frac{3\tm \ba}{\bp}-6\chi\ba \cos\bi\left(\frac{\tm}{\bp}\right)^{3/2}-\frac{3\tm^2\ba}{4 \bp^2}\left(10-\ba^2-\bb^2\right)+\frac{3\tm^2\ba\chi^2}{4\bp^2(1+D)}\left(5\cos^2\bi-1\right)\; ,\\
	\frac{\td \bi}{\td u}&=0\; ,\\
	\frac{\td\bO}{\td u} &=2\chi\left(\frac{\tm}{\bp}\right)^{3/2}-\frac{3\tm^2\chi^2\cos \bi}{2\bp^2(1+D)}\; .
	\end{split}
	\end{equation}
	The corresponding relations for $\{\bo,\be\}$ read
	\begin{equation}
	\label{omega_e_general}
	\begin{split}
	\frac{\td \bo}{\td u}& = \frac{3\tm}{\bp}-6\chi \cos\bi\left(\frac{\tm}{\bp}\right)^{3/2}-\frac{3\tm^2(10-\be^2)}{4\bp^2}+\frac{3\tm^2\chi^2(5\cos^2\bi-1)}{4\bp^2(1+D)}\; ,\\
	\frac{\td\be}{\td u}& = 0\; .\\
	\end{split}
	\end{equation}
	The expressions for the Kerr metric are obtained by setting $D=0$ in the above equations. 
	Note that terms of order $\mathcal{O}(\varepsilon^{n+1})$ in the metric correspond to $\mathcal{O}(\varepsilon^{n})$ order in the equations for the secular shifts~(\ref{sec_general}).
	In particular, in the Newtonian approximation, the right-hand side of Eq.~(\ref{sec_general}) is identically zero, so that there are no shifts in any of the orbital parameters.
	The leading-order PN corrections of $\mathcal{O}(\varepsilon)$ lead to a variation of $\alpha$ and $\beta$, which correspond to the standard pericenter precession as in GR. 
	The Lense-Thirring (or frame-dragging) effect is due to the term $\mathcal{O}(\varepsilon^{3/2})$, corresponding to 1.5PN order, in the last equation of Eq.~(\ref{sec_general}), which results in a variation of $\Omega$. Similar terms also enter the corrections to the shifts of $\alpha$ and $\beta$.
	The higher-order Schwarzschild corrections at 2PN order in the variation of $\alpha$ and $\beta$ (the third term on the right-hand side of Eq.~(\ref{sec_general})), are unaffected by the modification of gravity in this case.  Crucially however, quadrupole corrections proportional to $\chi^2$ now get corrected by the factor $(1+D)^{-1}$.
 As expected, the secular variation of the orbit parameters remains unchanged up to the Lense-Thirring terms when compared to Kerr, while the quadrupole terms are modified.

	%%%%%%%%%%%%%%%%%%%%%%%%%%%%%%%%%%%%%%%%%%%%%%%%%%%%%%%%%%%%%%%%%%%
	\subsection{A noncircular Schwarzschild deformation}
	\label{case_infty}
	Let us now consider the limit of an infinite disformal parameter, i.e. $D\to\infty$, while at the same time keeping the physical spin of the black hole $\tilde{a}=a\sqrt{1+D}$ finite. This implies $a\to 0$ but as we discussed above, the observable quantity is $\tilde{a}$ rather than $a$.
	This limit applied to the disformed metric~(\ref{metric}) yields the following finite line element:
	\begin{equation*}
	\td \tilde{s}^2_{\text{NCS}} = -\left(1-\frac{2\tm}{r}\right)\td t^2 + \sqrt{\frac{2 r}{\tm}}\td t\td r-\frac{4\chi \tm^2 \sin^2\theta}{r}\td t \td \vp-\frac{r}{2 \tm}\td r^2 + r^2 \td \theta^2 + r^2\sin^2\theta\left(1 + \frac{2\chi^2\tm^3\sin^2\theta}{r^3}\right)\td\vp^2\; .
	\end{equation*} 
	We can put the metric in a ``Schwarzschild-like" form,
	by trading the $(tr)$ term for an $(r\vp)$ term through the coordinate change~(\ref{time_redef}), assuming the limit $D\to \infty$ and $a\to 0$,
	\begin{equation}
	\label{metric_infty}
	\td \tilde{s}_{\text{NCS}}^2 = -\left(1-\frac{2\tm}{r}\right)\left(\td t + \frac{2\chi\tm^2\sin^2\theta}{r-2\tm}\td\vp\right)^2 + \left(1-\frac{2\tm}{r}\right)^{-1}\left(\td r -\sqrt{\frac{2\tm^3}{ r}}\chi\sin^2\theta\td\vp\right)^2 + r^2\left(\td\theta^2 + \sin^2\theta \td\vp^2\right)\; .
	\end{equation}
The metric is noncircular and given its familiar form (for $\chi=0$ the line element~(\ref{metric_infty}) corresponds to the Schwarzschild metric) we call the metric noncircular Schwarzschild (NCS) deformation. Despite the name, the properties of the above metric are quite different from the static GR case. For a start the metric is stationary and spinning :
	from the $(t\vp)$ term we can read off the value of the spin $\chi$, when compared to an asymptotic expansion of the Kerr spacetime. Also note that $r=2\tilde{M}$ is not a horizon location, but rather the static limit ergosurface. The horizon is a null surface situated in the interior of this ergosurface~\cite{Anson:2020trg}. A necessary condition for this disformed black hole to have a smooth outer horizon is $\chi<1/2$; see Ref.~\cite{Anson:2020trg}. The metric is everywhere regular apart from $r=0$ which is now the location of the black hole singularity. 
	As in the case of generic $D$, the nonzero contribution to the Ricci tensor starts at $\varepsilon^3$ order. It can be shown that this metric is a solution of a particular class of scalar-tensor theories belonging to DHOST Ia; see Appendix~\ref{app:disformed}.
	
	Similarly to the generic case considered above, we change coordinates to those that are harmonic for the Kerr metric,  and expand in $\varepsilon$ to obtain
	\begin{equation}
	\label{harm_infinity}
	\begin{split}
	\tilde{g}_{00}^{{(\text{NCS})}}&= -1+\frac{2\tm}{r}-\frac{2\tm^2}{r^2}+\frac{2\tm^3}{r^3} +\frac{\tm^3\chi^2}{r^3}\left[1-\left(\textbf{n}\cdot\textbf{s}\right)^2\right]+\mathcal{O}\left(\varepsilon^4\right)\; ,\\
	\tg_{0j}^{{(\text{NCS})}}&=\frac{2\tm^2\chi}{r^2}\left(\textbf{n}\times\textbf{s}\right)_j+ \mathcal{O}\left(\varepsilon^3\right)\; ,\\
	\tg_{ij}^{{(\text{NCS})}}&=\left[1+\frac{2\tm}{r}+\frac{\tm^2}{r^2}\left(1-\chi^2\right)\right]\delta_{ij}+2\sqrt{2}\chi\left(\textbf{n}\times\textbf{s}\right)_{(i}n_{j)}\left(\frac{\tm}{r}\right)^{3/2}+\frac{\tm^2 }{r^2}n_i n_j \\ &\qquad +\frac{\tm^2\chi^2}{r^2}\left[2 n_i n_j+s_i\varepsilon s_j-2s_{(i}n_{j)}(\textbf{n}\cdot\textbf{s})\right]+\mathcal{O}\left(\varepsilon^{5/2}\right)\; .
	\end{split}
	\end{equation}
	where now $r$ is the radial harmonic coordinate relevant for the PN expansion. 
	This metric is almost identical to the $D\to\infty$ limit of Eq.~(\ref{harm_no_lim}), the only difference being the term $\sim\mathcal{O}(\varepsilon^{3/2})$ in the $\tg^{(\text{NCS})}_{ij}$ components. This term comes from the $\tg_{r\vp}$ component of Eq.~(\ref{metric_infty}), meaning that the metric is already noncircular at this PN order (unlike in the case of generic $D$ discussed above). This implies, in particular, that the NCS metric cannot be matched to the BI metric~(\ref{BI_BL}).
	
	We now perform the two-timescale analysis described in~Sec.~\ref{sec:method}. We obtain
	\begin{equation}
	\label{sec_infinity}
	\begin{split}
	\frac{\td \bp}{\td u}& = 0\; ,\\
	\frac{\td\ba}{\td u}& = -\frac{3\tm \bb}{\bp}+6\chi\bb \cos\bi\left(\frac{\tm}{\bp}\right)^{3/2}+\frac{3\tm^2\bb}{4 \bp^2}\left(10-\ba^2-\bb^2\right)\; ,\\
	\frac{\td\bb}{\td u}& = \frac{3\tm \ba}{\bp}-6\chi\ba \cos\bi\left(\frac{\tm}{\bp}\right)^{3/2}-\frac{3\tm^2\ba}{4 \bp^2}\left(10-\ba^2-\bb^2\right)\; ,\\
	\frac{\td \bi}{\td u}&=0\; ,\\
	\frac{\td\bO}{\td u} &=2\chi\left(\frac{\tm}{\bp}\right)^{3/2}\; .
	\end{split}
	\end{equation}
	The above expressions can be alternatively found by taking the limit $D\to\infty$ in in Eq.~(\ref{sec_general}). Hence, the noncircular terms in the spatial components of Eq.~(\ref{harm_infinity}) do not influence the secular shifts at this PN order, as their effect averages to 0 over an orbital period.
	As we can see by comparing Eqs.~(\ref{sec_infinity}) and~(\ref{sec_general}) where we set $D=0$, the variation of the orbit elements are modified at 2PN order, while the 1PN and Lense-Thirring terms remain the same.
	The difference appearing at 2PN order is in the quadrupole terms: for the case under consideration they do not appear at this order, while in the Kerr case they are present.
	In terms of $\bo$, we obtain to 2PN order,
	\begin{equation}
	\frac{\td \bo}{\td u}=\frac{3\tm}{\bp}-6\chi\cos\bi\left(\frac{\tm}{\bp}\right)^{3/2}-\frac{3\tm^2(10-\be^2)}{4 \bp^2}\; .
	\end{equation}
	In comparison to the Kerr case, in the above expression the quadrupole term is absent, while other terms are the same as in the Kerr case, namely, the 1PN and 2PN Schwarzschild corrections and the Lense-Thirring term are recovered.

	%%%%%%%%%%%%%%%%%%%%%%%%%%%%%%%%%%%%%%%%%%%%%%%%%%%%%%%%%%%%
	\subsection{Quasi-Weyl metric}
	\label{case_-1}
	Now we consider the limit $D\to -1$ in the disformed metric~(\ref{metric}). 
We also assume that $\tilde{a}\to 0$, while $a$ remains finite. This is done in order for the asymptotic expansion~(\ref{metric_as}) to make sense, as otherwise the  terms proportional to $(1+D)^{-1}$ in the expansion become infinite.
	Although the physical rotation parameter $\tilde{a}$ is zero, this limit does not result in the Schwarzschild metric, which would correspond to $a\to 0$ instead. 
	However, the nondiagonal term $\tilde{g}_{t\vp}$ drops out and we obtain the following simple line element:
	\begin{equation}
	\label{formal_1}
	\td\tilde{s}_{\text{QW}}^2 = -\left(1-\frac{2\tilde{M}r}{\rho^2}\right)\td t^2 
	+  \frac{\rho^2 }{r^2 + a^2}\td r^2 + 2 \sqrt{\frac{2\tilde{M}r}{r^2 + a^2}}\td t\td r + \rho^2 \td \theta^2+ \left(r^2 + a^2\right)\sin^2\theta\td\vp^2\; .
	\end{equation}
	Now in the absence of the $\tilde{g}_{tr}$ term the above line element would be a Weyl metric, in essence a static and axially symmetric circular metric. The metric is however clearly not static; it is rather stationary and noncircular. At a loss of a better name we call this metric a quasi-Weyl (QW) metric; in essence it is a simple noncircular metric with no frame-dragging term in the $\vp$ direction. It has a ring singularity at $\rho=0$, and an ergosurface at $\tilde{g}_{tt}=0$, which interestingly is both  the static and stationary observer limit due to the absence of frame dragging. The event horizon on the other hand is in the interior of this surface, as found generically in Ref.~\cite{Anson:2020trg}. Despite the seemingly singular nature of the limit, it can be shown that this metric is a solution of a particular class of scalar-tensor theories belonging to DHOST Ia; see Appendix~\ref{app:disformed}.
	
	This case is in some sense the opposite of the $D\to\infty$ case studied above, where the Lense-Thirring effect is present but there are no quadrupole terms.
	If we set $a=0$, Eq~(\ref{formal_1}) corresponds to the Schwarzschild solution written in the Gullstrand–Painlev\'e coordinates. 
	After the coordinate change (\ref{time_redef}), the metric reads
	\begin{equation}
	\begin{split}
	\label{metric_QW}
	\td\tilde{s}_{\text{QW}}^2 &= -\left(1-\frac{2\tilde{M}r}{\rho^2}\right)\td t^2 
	+ \frac{r^5(r-2\tm)+2a^2 r^3 \cos^2\theta(r-3\tm)+a^4\cos^4\theta(r-2\tm)^2}{\rho^2(r-2\tm)^2(r^2+a^2)}\td r^2 \\
	&\qquad -\frac{4 a^2 \sqrt{2\tm^3 r}\cos^2\theta}{\rho^2(r-2\tm)\sqrt{a^2+r^2}}\td t\td r + \rho^2 \td \theta^2+ \left(r^2 + a^2\right)\sin^2\theta\td\vp^2\; .
	\end{split}
	\end{equation}
The 2PN expression of the above metric can be obtained from Eq.~(\ref{metric_as}) by defining $\chi_1=a/\tm$, setting $\chi = \chi_1\sqrt{1+D}$ and taking the limit $D\to-1$. Note that we also assume $\chi_1\sim\mathcal{O}(1)$. 
At this PN order, the quasi-Weyl metric can also be matched to the BI metric~(\ref{BI_BL}), with the BI parameters defined as $q=-\chi_1^2$ and $a_0 = \chi_1^2/12$.
The resulting 2PN metric does not contain the Lense-Thirring term, which could be anticipated, since the physical rotation parameter in this case is zero. 
Meanwhile the metric still contains a free quadrupole parameter.
As in the two previous cases, the Ricci tensor for the metric~(\ref{metric_QW}) is nonzero only at $\varepsilon^3$  order.
	
	Similarly to the generic case described above, we change to Kerr harmonic coordinates and calculate the secular variations of orbital parameters, 
	following the method of Sec.~\ref{sec:method}.
	We obtain the following results up to 2PN order:
	\begin{equation}
	\label{sec_formal}
	\begin{split}
	\frac{\td \bp}{\td u}& = 0\; ,\\
	\frac{\td\ba}{\td u}& = -\frac{3\tm \bb}{\bp}+\frac{3\tm^2\bb}{4 \bp^2}\left(10-\ba^2-\bb^2\right)-\frac{3\tm^2\bb\chi_1^2}{4\bp^2}\left(5\cos^2\bi-1\right)\; ,\\
	\frac{\td\bb}{\td u}& = \frac{3\tm \ba}{\bp}-\frac{3\tm^2\ba}{4 \bp^2}\left(10-\ba^2-\bb^2\right)+\frac{3\tm^2\ba\chi_1^2}{4\bp^2}\left(5\cos^2\bi-1\right)\; ,\\
	\frac{\td \bi}{\td u}&=0\; ,\\
	\frac{\td\bO}{\td u} &=-\frac{3\tm^2\chi_1^2\cos \bi}{2\bp^2}\; ,
	\end{split}
	\end{equation}
	where $\chi_1$ is a finite quantity, while $\chi=\tilde{a}/\tm$ introduced earlier is zero in this case.
	The above expressions can also be obtained from (\ref{sec_general}) by substituting $\chi = \chi_1\sqrt{1+D}$ as explained above.
	As one can see from~(\ref{sec_formal}), the Lense-Thirring terms drop out in this limit, which is consistent with the absence of a $(t\vp)$ term in the metric. 
	However, the quadrupole terms appear at 2PN order, similarly to the Kerr case. 
	While the structure of these terms is the same as for Kerr, the free parameter $\chi_1$ entering the quadrupole terms is not related to the black hole spin, which is zero in the quasi-Weyl case.  
This is to be compared to the Kerr case, for which the corresponding quadrupole term is constrained by the no-hair theorem.
	Indeed, for the Kerr metric with parameters $\{\tm,\tilde{a}\}$, the quadrupole reads, in terms of these, $Q =- \tm^3 \chi^2$. 
	Therefore if the no-hair theorem was valid in the case of quasi-Weyl, it would imply $Q=0$, since $\chi=0$.
	The metric~(\ref{formal_1}) instead contains an arbitrary quadrupole $Q_1=-\tm^3 \chi_1^2$, i.e. the no-hair theorem does not hold.

	%%%%%%%%%%%%%%%%%%%%%%%%%%%%%%%%%%%%%%%%%%%%%%%%%%%%%%%%%%%%%%%%%%%%%%%%
	\subsection{Enhanced Kerr disformation }
	%%%%%%%%%%%%%%%%%%%%%%%%%%%%%%%%%%%%%%%%%%%%%%%%%%%%%%%%%%%%%%%%%%%%%%%%%
	\label{case_1+D_small}
	Finally, for the two last variants of the Kerr disformation we examine the situation when $(1+D)$ is small but finite.
	As we saw above, the generic values of $D$ result in a rather mild effect on the secular shifts, i.e. only quadrupole terms in Eq.~(\ref{sec_general}) are modified. 
	The limit $D\to -1$ (quasi-Weyl) yields stronger modifications, since $\chi_{\text{QW}}=0$, and thus the frame-dragging terms are also modified with respect to Kerr. 
	In contrast to the quasi-Weyl case, here we assume that $D$ has a small finite offset from $-1$, so that the physical spin remains finite, while the corrections to the Kerr metric are enhanced with respect to the generic case. 
	Indeed, if we take $1+D \sim \varepsilon$, the terms proportional to $(1+D)^{-1}$ in the metric expansion~(\ref{metric_as})  become one order lower in $\varepsilon$. 
	More precisely, we assume the following form for the constant disformal factor:
	\begin{equation}
	\label{assumptions_Dneg}
	D = -1 + \frac{\chi^2}{\lambda} \varepsilon\; ,\qquad \{\lambda,\chi\}\sim\mathcal{O}\left(1\right)\; ,
	\end{equation}
	where the factor $\chi^2/\lambda$ is chosen for convenience and we introduced a new parameter $\lambda$ here.
	Similarly to the case of generic $D$ discussed above, we perform the coordinate transformation~(\ref{time_redef}) in the disformal metric (\ref{metric}), 
	and expand the line element to $\mathcal{O}(\varepsilon^3)$, assuming $\td x^i \sim \sqrt{\varepsilon}\, \td t$. 
	The result is
	\begin{equation}
	\label{metlimBL}
	\begin{split}
	\td \tilde{s}_{\text{EKD}}^2&\simeq -\left(1 - \frac{2A \varepsilon}{r} + \frac{2A^3 \varepsilon^2\lambda}{r^3}\cos^2\theta-\frac{2 A^5 \varepsilon^3 \la^2 \cos^4 \theta}{r^5}\right)\td t^2 \\ &+ \left(1 + \frac{A \varepsilon(2r - A \la \sin^2\theta)}{r^2}+\frac{A^2 \varepsilon^2(4r^2 - 2Ar\la + A^2\la^2\sin^2\theta)}{r^4}\right)\td r^2\\ & + r^2 \left(1 +\frac{ \varepsilon\lambda A^2}{r^2} \cos^2\theta\right)\td \theta^2 + r^2\sin^2\theta\left(1 + \frac{\varepsilon\lambda A^2}{r^2}\right)\td\vp^2-\frac{4\sqrt{2}A^{7/2}\varepsilon^{5/2}\la \cos^2\theta}{r^{7/2}}\td t \td r-\frac{4 A^2 \varepsilon^2 \chi\sin^2\theta}{r}\td t\td \vp\; .
	\end{split}
	\end{equation}
	Note that the above expression cannot be obtained by substituting Eq.~(\ref{assumptions_Dneg}) in the asymptotic expansion~(\ref{metric_as}). 
	This is because in Eq.~(\ref{metric_as}) we neglected, in particular, terms of the form $\sim (1+D)^{-2}\mathcal{O}(\varepsilon^5)$  in the $(tt)$ component of the metric, which become $\sim \mathcal{O}(\varepsilon^3)$ for values of the disformal parameter~(\ref{assumptions_Dneg}).
	Similarly, terms $\sim (1+D)^{-1}\mathcal{O}(\varepsilon^{7/2})$ in the $(tr)$ components and $\sim (1+D)^{-1}\mathcal{O}(\varepsilon^{3})$ or $\sim (1+D)^{-2}\mathcal{O}(\varepsilon^{4})$ in spatial components were neglected in Eq.~(\ref{metric_as}). However, they become important for the case considered here. 
	
The metric~(\ref{metlimBL}) is noncircular and its Ricci curvature is nonzero already at $\varepsilon^2$ order, which is one order lower than all other cases considered above.
As one of the consequences of noncircularity, the enhanced Kerr disformation (EKD) metric cannot be matched to the BI metric~(\ref{BI_BL}) at this order.
	Also, due to noncircularity in the generic case~(\ref{metric_as}), the off-diagonal term $(tr)$  cannot be eliminated in Eq.~(\ref{metlimBL}), since $\tilde{g}_{tt}$ depends on $\theta$ now. 
	To recover the asymptotic Kerr metric at $\mathcal{O}(\varepsilon^3)$ one replaces $\la=\chi^2\varepsilon$ in Eq.~(\ref{metlimBL}), which corresponds to setting $D=0$. 
	By inverting the relations (\ref{Kerr_harm}) to the right order, the asymptotic expansion~(\ref{metlimBL}) can be written in the Kerr harmonic coordinates,
	\begin{equation}
	\begin{split}
	\label{metric_harm}
	\tg^{(-1)}_{00}&= -1 + \frac{2\tm}{r}-\frac{2\tm^2}{r^2}\left[1+ \frac{\tilde{A}\left(\textbf{n}\cdot\textbf{s}\right)^2}{r}\right]+\frac{\tm^3}{r^3}\left[2 + \left(1-(\textbf{n}\cdot\textbf{s})^2\right)\chi^2 +\frac{6\tilde{A}(\textbf{n}\cdot\textbf{s})^2}{r}+\frac{2\tilde{A}^2(\textbf{n}\cdot\textbf{s})^4}{r^2}\right]+\mathcal{O}\left(\varepsilon^4\right)\; ,\\
	\tg^{(-1)}_{0j}&= \frac{2\tm^2\chi}{r^2}\left(\textbf{n}\wedge\textbf{s}\right)_j-\frac{2\sqrt{2}\tilde{A}\tm^{5/2}(\textbf{n}\cdot\textbf{s})}{r^{7/2}}n_j+\mathcal{O}\left(\varepsilon^3\right)\; ,\\
	\tg^{(-1)}_{ij}&=\left[1+\frac{\tm}{r}\left(2+\frac{\tilde{A}}{r}\right)+\frac{\tm^2}{r^2}\left(1-\chi^2\right)\right]\delta_{ij}-\frac{2\tilde{A}\tilde{M}}{r^2}n_in_j +\frac{\tm^2}{r^2}\left[1+ 2\chi^2-\frac{2\tilde{A}(\textbf{n}\cdot\textbf{s})^2}{r}+\frac{\tilde{A}^2}{r^2}\left(1-(\textbf{n}\cdot\textbf{s})^2\right)\right]n_i n_j\\
	& \qquad+ \frac{\tilde{A}\tm}{r^2}\left[1-\frac{\tm \chi^2}{\tilde{A}}\right]\left[2s_{(i}n_{j)}(\textbf{n}\cdot\textbf{s})-s_i s_j\right]+\mathcal{O}\left(\varepsilon^{5/2}\right)\; ,
	\end{split}
	\end{equation}
	where $\tilde{A}= \la A$. The asymptotic expansion of the Kerr metric in harmonic coordinates is recovered by setting $\tilde{A}\to \tm \chi^2$ and keeping terms up to 2PN order (one can check that the Kerr metric is indeed recovered by comparing to Ref.~\cite{Will:2016pgm} for instance).
	
	We now apply the osculating orbit method starting with the metric~(\ref{metric_harm}). 
	The metric is written up to 2PN order, and thus we can calculate the secular shifts up to this order. 
	The final expressions are quite heavy, and can be found in Appendix~\ref{app:secular}. 
	Setting $\la=\chi^2 \varepsilon$, our result up to 2PN order coincides with~Ref.\cite{Will:2016pgm} for the Kerr spacetime.
	Expressing the result in terms of $\{\be,\bo\}$, we can see that the secular variation of the parameters $\{\bp,\bi,\be\}$ is of 2PN order. 
	This is to be compared with the Kerr case, for which the effect is only of 3PN order. 
	This naively suggests possible strong secular effects in the case of the disformal metric, meaning that the parameters $\{\bp,\bi,\be\}$ could change considerably over a long period of time, when compared to the Kerr predictions.
	However, this is not the case, because the corresponding corrections in Eq.~(\ref{sec_pert}) average to zero. 
	Indeed, using the fact that the pericenter angle $\bo$ varies over a shorter timescale than the other parameters (the first correction appears at 1PN order), 
	we can average over $\bo$ in the system~(\ref{sec_pert}) \cite{Will:2016pgm}.
	After this average, there remains no secular variation of $\{\bp,\bi,\be\}$ at this order (see Appendix \ref{app:secular} for details)
		
	Using the expressions that we obtained for the variations of $\ba$ and $\bb$ in Eq.~(\ref{sec_pert}), one can derive the corresponding formula for the secular variation of $\bo$:
	\begin{equation}
	\frac{\td \bo}{\td u} = \frac{3\tm}{\bp}\left[1+\frac{\la}{4(1-\be^2)}\left(5\cos^2\bi -1\right)\right]-6\chi \cos\bi \left(\frac{\tm}{\bp}\right)^{3/2}+\mathcal{O}\left(\varepsilon^2\right)\; .
	\end{equation}
	When combined with the variation of $\bO$ and after multiplication by $2\pi$, one obtains
	\begin{equation}
	\label{leading_pericenter}
	\Delta \bar{\varpi}\equiv \Delta\bo+\cos\bi\,\Delta\bO=\frac{6\pi \tm}{\bp}\left[1+\frac{\la}{4(1-\be^2)}\left(3\cos^2\bi -1\right)\right] + \mathcal{O}\left(\varepsilon^{3/2}\right)\; ,
	\end{equation}
where $\bar{\varpi}$ is the precession of the pericenter relative to the fixed reference direction (the $(Ox)$ axis in our case; see Fig.~\ref{angle_def}).
	As expected from the asymptotic expression~(\ref{asymptotic}), there are leading-order corrections to the secular pericenter shift. In Appendix~\ref{app:textbook}, we derive this leading term using the standard textbook method in the equatorial plane. One can check that the two methods are compatible by setting $\bi=0$ in Eq.~(\ref{leading_pericenter}), which corresponds to an orbit in the equatorial plane of the black hole. While the leading-order terms are the same, one must be careful when comparing the higher-order corrections of the different methods, as explained in Ref.~\cite{Tucker:2018rgy}.
	\subsection{ {\it{e}\,}nhanced Kerr disformation}
	\label{case_1+D_small2}
	Here we examine a case  of small and finite deviation of $D$ from $-1$, similar to the previous case.  
	The difference is in the value of the offset; here we study larger (but still small) deviations of order $\sqrt{\varepsilon}$:
		\begin{equation}
		\label{D_sqrt}
		D=-1 + \frac{\chi^2}{\la_2}\sqrt{\varepsilon}\; ,\qquad\la_2\sim\mathcal{O}(1)\; .
		\end{equation}
		Since the offset is larger than in the previous case, one expects that the modified gravity effects are smaller than those for EKD, while still larger than in the generic case. The metric in this limit can be obtained by replacing the relation~(\ref{D_sqrt}) in the line element~(\ref{harm_no_lim}).
		Unlike in the EKD case, the metric is circular at 2PN order, and we have $R_{\mu\nu}\sim\varepsilon^{5/2}$. The resulting metric cannot be matched to the BI metric~(\ref{BI_BL}),  since by replacing the disformal parameter according to Eq.~(\ref{D_sqrt}), one introduces fractional powers of the mass in the metric.
		The secular variation of orbital parameters reads
		\begin{equation}
		\label{sec_sqrt}
		\begin{split}
		\frac{\td \bp}{\td u}& = 0\; ,\\
		\frac{\td\ba}{\td u}& = -\frac{3\tm \bb}{\bp}+\frac{3\bb}{4}\left(8\chi\cos \bi-\frac{\la_2(5\cos^2\bi-1)}{\sqrt{1-\ba^2-\bb^2}}\right)\left(\frac{\tm}{\bp}\right)^{3/2}+\frac{3\tm^2\bb}{4 \bp^2}\left(10-\ba^2-\bb^2\right)\; ,\\
		\frac{\td\bb}{\td u}& = \frac{3\tm \ba}{\bp}-\frac{3\ba}{4}\left(8\chi\cos \bi-\frac{\la_2(5\cos^2\bi-1)}{\sqrt{1-\ba^2-\bb^2}}\right)\left(\frac{\tm}{\bp}\right)^{3/2}-\frac{3\tm^2\ba}{4 \bp^2}\left(10-\ba^2-\bb^2\right)\; ,\\
		\frac{\td \bi}{\td u}&=0\; ,\\
		\frac{\td\bO}{\td u} &=2\left(\chi - \frac{3\la_2\cos\bi}{4\sqrt{1-\ba^2-\bb^2}}\right)\left(\frac{\tm}{\bp}\right)^{3/2}\; .
		\end{split}
		\end{equation}
The above secular variations can be obtained by replacing Eq.~(\ref{D_sqrt}) in the expressions~(\ref{sec_general}) for the generic case. 
As one can see from the above expressions, in this case the Kerr quadrupole terms drop out, similarly to the NCS case ($D\to\infty$). 
However for the {\it{e}}KD metric corrections appear at lower order, at the level of the Lense-Thirring terms.
This happens because the quadrupole (2PN) corrections of the generic case become 1.5PN-order corrections here, due to the presence of the small value of $(1+D)$ given in Eq.~(\ref{D_sqrt}). The 1PN order is not modified for the {\it{e}}KD case, unlike in the EKD case considered above.

Note that it is possible to consider deformations of the form $1+D\sim \varepsilon^j$ in Eq.(\ref{D_sqrt}), where $j\leq 1$. The case $j=1$ was discussed in section~\ref{case_1+D_small}, while $j>1$ would be ruled out from the Schwarzschild precession measurement by the GRAVITY Collaboration~\cite{Abuter:2018drb}, since the leading GR term is modified in this case. We chose the particular case $j=1/2$ for simplicity, since for other powers $j<1$, the term proportional to $\la_2$ in Eq.~(\ref{sec_sqrt}) would introduce a  fractional power $2-j$ of the mass either between the leading pericenter and Lense-Thirring terms (for $j>1/2$) or between the Lense-Thirring and quadrupole contributions (for $j<1/2$). 
	%%%%%%%%%%%%%%%%%%%%%%%%%%%%%%%%%%%%%%%%%%%%%%%%%%%%%%%%%%%%%%%%%%
	\section{Summary, observational constraints and predictions}
	%%%%%%%%%%%%%%%%%%%%%%%%%%%%%%%%%%%%%%%%%%%%%%%%%%%%%%%%%%%%%%%%%%
	\label{sec:conclusion}
	In this work, we have studied the 2PN motion of stars around a disformed Kerr black hole, and compared the effects to those predicted by the Kerr metric. 
	For the case of a generic $D$, we found that the modification of gravity appears at the quadrupole level, leading to a violation of the no-hair theorem.
	We also examined particular limiting cases of $D$, which were motivated by the search for simpler line elements and/or larger deviations from GR that can be tested experimentally.

	For a start, the NCS metric obtained in the limit $D\to\infty$ and studied in Sec.~\ref{case_infty}, constitutes an interesting hybrid between the Schwarzschild and Kerr spacetimes. 
	While the $(t-r)$ sector of the metric is identical to the Schwarzschild spacetime, there remains a $(t\vp)$ term which introduces Lense-Thirring corrections to the orbit of stars, see Eq.~(\ref{metric_infty}). 
	Interestingly, the quadrupole corrections that arise for the Kerr spacetime are not present in this case. The NCS metric is a simple example of a noncircular metric, and its study could be helpful in understanding the effects of noncircularity. 
	In regard to observational predictions of the secular variation of orbital parameters, this case is quite similar to the generic case of $D$.
	In spite of the 1.5PN corrections in the metric (coming from the $\tilde{g}_{r\vp}$ term in Eq.~(\ref{metric_infty})),
	the secular variation of orbit elements is modified at 2PN order, while the 1PN and Lense-Thirring terms are the same  as in the Kerr case~(\ref{sec_infinity}).
	
	A second interesting limit of Kerr deformations arises when $(1+D)$ becomes small. In this case, as can be guessed e.g. from Eq.~(\ref{asymptotic}), 
	the corrections to the Kerr metric become more important than those for generic $D$, due to the terms containing $(1+D)$ in the denominator.
	We examined separately two cases: the exact limit $D\to -1$ and the case when $(1+D)$ is small but finite.
	In the former case we took the formal limit $D\to-1$ and $\tilde{a}\to0$ while at the same time keeping $a$ finite, and obtained the quasi-Weyl metric, see Sec.~\ref{case_-1}.
	Note that taking the limit of vanishing $\tilde{a}$ is needed in order to avoid the divergence of the metric when $D\to -1$.
	Since the observed spin is zero in this case, $\tilde{a}=0$, the Lense-Thirring effects are absent in the secular shifts given by Eq.~(\ref{sec_formal}). 
	It is interesting to note that while the quadrupole terms are present in the expressions for the variation of orbital parameters (the terms proportional to $\chi_1^2$ in Eq.~(\ref{sec_formal})), the parameter $\chi_1$ appearing in these terms is not related to the observed spin of the black hole. 
	
	Finally, we examined the case where $(1+D)$ is small but finite, while at the same time keeping a finite observable spin~$\tilde{a}$.
	In Sec.~\ref{case_1+D_small}, the case of the enhanced Kerr disformation, we assumed that $(1+D)\sim \varepsilon$. 
	In this case, the subdominant corrections of $\mathcal{O}(\varepsilon^{n+1})$ order (or higher) in the metric expansion containing $(1+D)$ in the denominator are comparable with $\mathcal{O}(\varepsilon^n)$ order terms of the Kerr metric.
	The asymptotic expansion in this case is more delicate, since we need to take into account higher-order terms in $\varepsilon$ in order to get secular shifts to 2PN order 
	(the result is given in Appendix~\ref{app:secular}). 
	The pericenter shift in this case receives  a leading-order correction due to the  modification of gravity, given by Eq.~(\ref{leading_pericenter}). 
	This case is therefore most interesting from the perspective of observing the effects of Kerr deformations.
	As a separate subcase we also considered a slightly larger offset of $(1+D)$ from zero, namely $1+D\sim\sqrt{\varepsilon}$, which we dubbed \emph{e}nhanced Kerr disformation.
	Such an assumption about the parameter of disformality results in the absence of quadrupole corrections in the secular variation of orbital parameters, while we obtained modifications of  the 1.5PN-order terms instead.
Table~\ref{tab:cases} summarizes the main results and particularities of different variants of the disformed Kerr metric, depending on the values of the disformal parameter.
	
\begin{table}[t]
		\setlength\extrarowheight{2pt}
		\centering
		\begin{tabular}{|m{3cm}||m{1.3cm}|m{2.1cm}|m{1.7cm}|m{2.2cm}|m{2.2cm}|}
			\hline
		 \multicolumn{6}{|c|}{Deviations from Kerr}
			\tabularnewline
			\hline
				\centering Value of D &\centering Metric &\centering Secular evolution  &\centering $R_{\mu\nu}$   &\centering Circularity at 2PN order&\centering BI form at 2PN order
			\tabularnewline
			\hhline{|=#=|=|=|=|=|}
			\centering Generic $D$ &  \centering 2PN & \centering2PN & \centering$\mathcal{O}(\varepsilon^3)$ & \centering\cmark & \centering\cmark
			\tabularnewline
			\hline
			\centering NCS $(D\to\infty)$ &  \centering 1.5PN & \centering2PN & \centering$\mathcal{O}(\varepsilon^3)$ & \centering\xmark& \centering\xmark
			\tabularnewline
			\hline
			\centering QW $(D\to -1)$ &  \centering 1.5PN & \centering1.5PN & \centering$\mathcal{O}(\varepsilon^3)$ & \centering\cmark& \centering\cmark
			\tabularnewline
			\hline
			\centering EKD $(D+1\sim\varepsilon)$ & \centering 1PN & \centering 1PN & \centering$\mathcal{O}(\varepsilon^2)$ & \centering\xmark& \centering\xmark
			\tabularnewline
			\hline
			\centering  \emph{e}KD $(D+1\sim\sqrt{\varepsilon})$ &  \centering 1.5PN & \centering 1.5PN & \centering$\mathcal{O}(\varepsilon^{5/2})$ & \centering\cmark& \centering \xmark
			\tabularnewline
			\hline
		\end{tabular}
		\caption{Summary of the different regimes considered for the disformal parameter $D$ and some properties of the resulting metrics at 2PN order. In the first column, we report the PN order at which the metric differs from Kerr in each case. The second column shows the PN order at which the predictions for the secular evolution of orbital parameters start to deviate from Kerr. The third column contains the order of the Ricci tensor. In the fourth column, we specify if the 2PN metric in each case is circular, and in the last column if the 2PN metric can be identified with the Butterworth-Ipser metric.}
		\label{tab:cases}
\end{table}

Observations of the star S2 in the center of our Galaxy provide an opportunity to test GR by measuring its redshift~\cite{Abuter:2018drb} and pericenter precession~\cite{Abuter:2020dou}. The redshift includes the Newtonian Doppler effect and relativistic corrections. 
The measured combination of the leading corrections, the gravitational redshift and relativistic transverse Doppler effect, was found to be in agreement with GR~\cite{Abuter:2018drb}. 
Note, however, that the gravitational redshift at this observational precision is due to the Newtonian potential in the $g_{tt}$ component of the metric.
This means that the current observations of the S2 star's redshift do not allow to test the Sgr~A* metric beyond Newtonian order. 
Taking into account that all the variants of the disformed Kerr metric agree with GR at this order, these observations do not put any constraints on the considered Kerr deformations.

The star S2 experiences a pericenter precession when orbiting around Sgr~A*. It was found to be in agreement with GR~\cite{Abuter:2020dou} 
with the accuracy $f_\text{SP} \simeq 1.1 \pm 0.2$, where $f_\text{SP}$ defines the ratio of the orbital pericenter precession (per period) of S2 to its GR value. 
For GR one has $f_\text{SP} =1$ while for Newtonian gravity $f_\text{SP} =0$.
Since the pericenter precession is a 1PN effect, the only case we can constrain using the observed pericenter precession~\cite{Abuter:2020dou} is the enhanced Kerr disformation studied in Sec.~\ref{case_1+D_small}\footnote{Note that since the pericenter precession is sensitive to 1PN-order terms, its measurement also allows to constrain spherically symmetric deformations of the Schwarzschild metric in DHOST theory. In particular Refs.~\cite{Zakharov:2019jio,Zakharov:2018awx} constrained a particular solution of Horndeski theory given in Ref.~\cite{Babichev:2017guv}.}. All other deformations give corrections to the orbital shifts at higher PN orders, see e.g. Table~\ref{tab:cases}, and therefore they automatically pass this observational test\footnote{Note also that for stronger Kerr deformations than EKD, for instance when $1+D\sim\varepsilon^{3/2}$, the correction to the pericenter shift is larger than the leading GR correction, which is already ruled out by the GRAVITY observations~\cite{Abuter:2020dou}.}. 
For the EKD metric, on the other hand, 
it is possible to constrain the disformal parameter $D$ using the pericenter precession of the star S2.
In order to stay within the experimental bounds of Ref.~\cite{Abuter:2020dou}, the correction to the Schwarzschild precession in Eq.~(\ref{leading_pericenter}) must satisfy 
\begin{equation}
\left|\frac{\la\left(3\cos^2\bi -1\right)}{4(1-\be^2)}\right|\lesssim 0.2\; .
\end{equation}
If we assume $|3\cos^2\bi -1|\sim1$, and replace the eccentricity $\be=0.87$ of the star S2, the inequality is saturated for $\la_0\sim 0.2$. Using the relation~(\ref{assumptions_Dneg}), we deduce a lower bound for the disformal parameter, $D\geq D_0$, which verifies:
	\begin{equation}
	D_0=-1+ \frac{\chi^2\ve_0}{\la_0}\; ,
	\end{equation}
where $\ve_0=\tm/A_0$, with $A_0$ being the semimajor axis of S2's orbit. 
Taking the deformation to be $D=D_0$ to maximize the effects of disformality,
we consider another star with $\ve\neq \ve_0$ in general.
The leading pericenter precession reads, from Eq.~(\ref{leading_pericenter}), 
\begin{equation}
\label{prediction}
\Delta \bar{\varpi}=\frac{6\pi \tm}{\bp}\left[1+\frac{\varepsilon\la_0}{\varepsilon_0}\frac{\left(3\cos^2\bi -1\right)}{4(1-\be^2)}\right] .
\end{equation}
Note the factor $\varepsilon\lambda_0/\varepsilon_0$ in the second term of the brackets, since $\varepsilon/\lambda = \varepsilon_0/\lambda_0$.
The above expression is correct as long as $\varepsilon$ stays in the range,
\begin{equation}
\label{range}
\ve^2\lesssim 10^{-3}\lesssim\sqrt{\ve}\; ,
\end{equation}
which implies that the perturbative expansion in $\varepsilon$ is valid\footnote{
To see this explicitly, we need to inspect the subleading terms in Eq.~(\ref{prediction}), which have the structure $\mathcal{O}\left((1+\lambda)\varepsilon^{3/2}\right) + \mathcal{O}\left((1+\lambda)^2\ve^2\right)$.
The first inequality in Eq.~(\ref{range}) comes from the requirement that $\mathcal{O}\left((1+\lambda)^2\ve^2\right)$ is subleading with respect to Eq.~(\ref{prediction}), i.e. $(\lambda\varepsilon)^2\lesssim (\lambda\varepsilon)$.
The second inequality comes from the comparison of Eq.~(\ref{prediction}) with $\mathcal{O}\left((1+\lambda)\varepsilon^{3/2}\right)$ for small $\lambda$: one ensures that the corrections are subdominant, resulting in~$\varepsilon^{3/2}\lesssim \lambda\varepsilon$. After replacing the values for $\la_0\sim 10^{-1}$ and $\ve_0\sim 10^{-4}$ we obtain Eq.~(\ref{range}).
}

	Future observations will determine the pericenter precession of other stars orbiting around Sgr A*.  If some of them have high eccentricities, the effect of modified gravity in the case of the EKD metric will be detected by the correction to the pericenter precession, as suggested by Eq.~(\ref{prediction}). 
	This is correct for a generic inclination angle~$\bar\iota$; however it is worth noting that for a specific value $\bi = \arccos(1/\sqrt{3})$, the contribution coming from the disformal metric vanishes completely. Note also that depending on the value of $\bi$, it is in principle possible to obtain a negative pericenter precession at leading order. This is a notable difference from the Kerr spacetime, as argued in Ref.~\cite{Bambhaniya:2021ybs}, where the authors showed that a negative precession can arise in the case of a naked singularity in the Johannsen-Psaltis spacetime \cite{Johannsen:2011dh}.

While all the variants of the Kerr deformations, besides the EKD metric, automatically satisfy the current observational bounds coming from the star S2, future experiments will be able to probe these Kerr deformations as well. Indeed, none of the deformations of Kerr presented in this paper verify the no-hair theorem. 
Therefore, future observations aiming at testing the no-hair theorem for the Kerr spacetime will probe all the deformations of Kerr. More precisely, the observation of high-eccentricity stars with short periods orbiting Sgr A* can in principle lead to the determination of the spin and quadrupole moment by measuring the secular variation of the nodal and inclination angles $\{\Omega,\iota\}$ \cite{Will:2007pp} (see also Ref.~\cite{Johannsen:2015mdd} for a review). Another promising method to test the no-hair theorem is to use pulsar timing, which could allow the determination of the spin and quadrupole moment of Sgr~A* if a binary pulsar orbiting closely enough to the black hole is discovered (see the review \cite{Johannsen:2015mdd} and references therein). The authors of Ref.~\cite{Christian:2015smg} calculated the second-order Shapiro delay for the BI metric. These results can be applied to the disformed Kerr metric in the cases where the line element can be put in the BI form (see table \ref{tab:cases}). The disformed metric in the generic case, as well as in the limits $D\to\infty$ and $D\to-1$ provide examples where the no-hair theorem is not verified, and hence are interesting counterexamples to the standard GR case.

\section*{Acknowledgements}
We are very happy to thank Vyacheslav Dokuchaev, Gilles Esposito-Far\`ese, François Larrouturou, Karim Noui and Alexander Zakharov for helpful discussions. C. C. thanks George Pappas and Nikolaos Stergioulas at the University of Thessaloniki for hospitality and discussions in the beginning stages of this project.
The work was supported by the CNRS/RFBR Cooperation program for 2018-2020 n. 1985 ``Modified gravity and black holes: consistent models and experimental signatures.''
	
		\appendix
	
	\section{Construction of the disformed metric and corresponding DHOST theories}
	\label{app:disformed}
	\subsection{Construction of the metric}
	The starting point of our construction is the stealth Kerr solution  \cite{Charmousis:2019vnf}, which belongs to the class Ia of DHOST theory.
	The metric in Boyer-Lindquist-like coordinates reads,
	\begin{align}
	\label{BL} g_{\mu\nu}\td x^\mu \td x^\nu &=
	-\left(1-\frac{2Mr}{\rho^2}\right)\td t^2
	-\frac{4Mar\sin^2\theta}{\rho^2}\td t\td\vp+
	\frac{\sin^2\theta}{\rho^2}\left[\left(r^2+a^2\right)^2-a^2\Delta
	\sin^2\theta\right]\td\vp^2 + \frac{\rho^2}{\Delta}\td r^2 + \rho^2
	\td \theta^2\; ,\\
	\label{phi} \phi &= q_0 \left[t + \int \frac{\sqrt{2M
			r(a^2+r^2)}}{\Delta}\td r\right]\;,
	\end{align}
	where $M$ represents the bare mass of the black hole, $a$ is the angular
	momentum per unit mass, $q_0$ is a positive constant, and $\{\rho,\Delta\}$ are defined in the main text after Eq.~(\ref{metric}).
	The nontrivial scalar $\phi$ defines a geodesic vector $\del_\mu\phi$, which is used to construct the disformed Kerr metric $\tilde{g}$ \cite{Anson:2020trg}:
	\begin{equation}
	\label{disformal} \tilde{g}_{\m\n} = g_{\m\n} - \frac{D}{q_0^2}\;
	\del_\mu\phi\,\del_\nu\phi\; ,
	\end{equation}
	where $D$ is a constant whose sign is not fixed \emph{a priori}. 
	After introducing the rescaled mass $\tilde{M} =M/(1+D)$, which is in fact the physical mass as measured by a distant observer, and rescaling the time coordinate as $t\to t/ \sqrt{1+D}$, we obtain our starting metric (\ref{metric}). Note that we have assumed $D+1>0$ to perform these redefinitions.
	Although $D$ is a given constant parametrizing a particular DHOST theory, it can be thought of as an additional parameter of the disformed metric in a more phenomenological approach.
	A nontrivial addition to the new metric is the term $\tilde{g}_{tr}$, which cannot in general be eliminated by a coordinate change without introducing other off-diagonal elements (see the discussion in Ref.~\cite{Anson:2020trg}). An exception is the static case $a=0$, for which it can be shown by an appropriate coordinate change that the metric is simply that of a Schwarzschild black hole of mass $\tm$  \cite{Babichev:2017lmw,Babichev:2018uiw}.
	
	\subsection{DHOST theories}
	Since the DHOST Ia class is stable under the disformal map \cite{Achour:2016rkg}, the disformed metric (\ref{metric}) is again a solution of a DHOST theory determined by the disformal parameter $D$. The shift-symmetric quadratic DHOST action \cite{Langlois:2015cwa,Crisostomi:2016czh} can be written in the following way:
	\begin{equation}
	\label{actionDHOST}
	S[g] = \frac{M_P^2}{2}\int\td^4 x \sqrt{-g}\left( f(X)R+K(X)-G_3(X)\square\phi + \sum_{i=1}^{5}A_i(X)\mathcal{L}_i\right)\; , 
	\end{equation}
	where $M_P^2 = (8\pi G)^{-1}$ is the reduced Planck mass,$X=g^{\mu\nu}\del_\mu\phi\del_\nu\phi$, and the $\mathcal{L}_i$ are given by
	\begin{equation}
	\label{L}
	\mathcal{L}_1=\p_{\m\n}\p^{\m\n}, \quad 
	\mathcal{L}_2=(\square\phi)^2, \quad 
	\mathcal{L}_3=\phi_{\mu\n}\p^\m\p^\n \square\phi, \quad
	\mathcal{L}_4= \phi_\mu\phi^\nu\phi^{\mu\alpha}\phi_{\nu\alpha}, \quad
	\mathcal{L}_5= \left(\phi_{\mu\n}\p^\m\p^\n\right)^2\; ,
	\end{equation}
	with $\phi_\mu\equiv \nabla_\mu\phi$ and $\phi_{\mu\nu}\equiv\nabla_\mu\nabla_\nu\phi$.
	In order for the theory to be degenerate, the functions $\{A_2,A_4,A_5\}$ are fixed by the choice of $\{f,A_1,A_3\}$. We will focus on the DHOST Ia class, which is related to the Horndeski theories by the disformal map. The functions in the Lagrangian are modified by the transformation~(\ref{disformal}), such that we have $\tilde{S}[\tg]=S[g]$. 
We couple the matter fields minimally in each case (either to $\tg_{\mu\nu}$ or $g_{\mu\nu}$), which ensures that we indeed have different theories. 
For a constant disformal parameter, the transformation rule for the Lagrangian functions \cite{Achour:2016rkg} is simplified, and we have
	\begin{equation}
	\label{function_disf}
	\begin{split}
	f &= \tilde{f}\sqrt{1+BX}\; ,\\
	A_1 &= \frac{\tilde{A}_1+B(1+BX)\tilde{f}}{(1+BX)^{3/2}}\; ,\\
	A_2&=\frac{\tilde{A}_2-B(1+BX)\tilde{f}}{(1+BX)^{3/2}}\; ,\\
	A_3 &=\frac{\tilde{A}_3 - 2 B(1+BX)\tilde{A}_2 - 4 B (1+BX)^3 \tilde{f}_X }{(1+BX)^{7/2}}\; ,\\
	A_4 &=\frac{\tilde{A}_4 - 2 B(1+BX)\tilde{A}_1 + 4 B (1+BX)^3 \tilde{f}_X }{(1+BX)^{7/2}}\; ,\\
	A_5 &=\frac{\tilde{A}_5+B(1+BX)\left[B(1+BX)(\tilde{A}_1+\tilde{A}_2)-(\tilde{A}_3+\tilde{A}_4)\right]}{(1+BX)^{11/2}}\; ,
	\end{split}
	\end{equation}
	where $B=-D/q_0^2$ and $\tilde{f}_X=\del \tilde{f}/\del X$. The functions of $\tilde{X}$ can be seen as functions of $X$ through the transformation
	\begin{equation}
	\label{Xtilde}
	\tilde{X}=\frac{X}{1+BX}\; .
	\end{equation}
	Since we start with a stealth configuration where $X=-q_0^2$, and we assume a constant disformal factor, it is clear that $\tilde{X}$ is again a constant on-shell. This means that the $\{\tilde{A}_4,\tilde{A}_5\}$ terms in the Lagrangian will not enter the equations of motion. Indeed, one has $\mathcal{L}_4\sim X^\mu X_\mu$ and $\mathcal{L}_5\sim (X^\mu \phi_\mu)^2$, so any variation of these terms will include at least one derivative of $X$ and hence vanish in the field equations. For this reason, we will be interested in the functions $\{\tilde{f},\tilde{A}_1,\tilde{A}_3\}$ in the following, as we impose $\tilde{A}_2=-\tilde{A}_1$ which is a condition for not having the Ostrogradsky ghost in DHOST Ia theories. Additionally, we will assume that the initial theory (with the stealth-Kerr solution), satisfies $A_1=A_2=G_3=0$~\cite{Charmousis:2019vnf}, and we also assume $K=0$. Then, inverting the relations~(\ref{function_disf}), we obtain:
	\begin{equation}
	\label{functions_general}
	\begin{split}
	\tilde{f}&=\frac{f}{\sqrt{1+BX}}\; ,\\
	\tilde{A}_1&=-B\sqrt{1+BX}f\; , \\
	\tilde{A}_3&= (1+BX)^{7/2}A_3 + 4 B(1+BX)^{5/2} f_X\; .
	\end{split}
	\end{equation}
	To write the action in terms of $\tilde{X}$, one uses the relation $1+BX=1/(1-B\tilde{X})$. In particular, we see that setting $\tilde{f}=\sqrt{1-B \tilde X}$ and $A_3=0$, gives us the particular Horndeski theory which admits the generic disformal metric  (\ref{metric}) as a solution (see also Ref.~\cite{Babichev:2017guv}). 
	
	As it is clear from the above discussion, any transformations involving finite nonzero $1+D>0$ are well defined. 
		This implies, in particular, that the metric for the case of generic $D$, as well as EKD and \emph{e}KD metrics are solutions of DHOST theories. 
		From the point of view of the action, the only difference is the choice of parameter $D$. 
		On the other hand, the two other cases, NCS and QW metrics, are obtained by taking the limit $D\to \infty$ and $D\to -1$. 
		Neither of these limits are well defined at the level of the action, as formulated above. Thus it is not guaranteed that the NCS or QW metrics are solutions of any finite nonsingular theory. However, as we show below, it is possible in both of these cases to formulate an action leading to the NCS and QW metrics as solutions.
	
	%%%%%%%%%%%%%%%%%%%%%%%%%%%%%%%%%%%%%%%%%%%%%%%%%%
	\subsection{Theory giving rise to the noncircular Schwarzschild metric}
	The NCS metric presented in Sec.~\ref{case_infty} is obtained by taking the limit $D\to \infty$.
		In order to make sense of the action we first redefine the scalar field as $\phi=q_0\psi/\sqrt{1+D}$, and the expression for the disformed metric~(\ref{disformal}) and scalar kinetic term become
	\begin{equation*}
	\begin{split}
	\tilde{g}_{\mu\nu}&=g_{\mu\nu}-\frac{D}{1+D} \del_\mu\psi\del_\nu\psi\; ,\\
	Y&= g^{\mu\nu}\del_\mu\psi\del_\nu\psi = \frac{D}{q_0^2}X\; .
	\end{split}
	\end{equation*}
	The resulting action in terms of $\psi$ and $\tilde{Y}=Y/(1-Y)$ reads, in the limit $D\to \infty$,
	\begin{equation*}
	S_{D\to\infty}[\tilde{g}] = \frac{M_P^2}{2}\int\td^4 x \sqrt{-\tilde{g}}\left( \sqrt{1+\tilde{Y}}f(\tilde{Y})\tilde{R} +\frac{f}{\sqrt{1+\tilde{Y}}}\left[\psi_{\mu\nu}\psi^{\mu\nu}-(\square\psi)^2\right] -\frac{4f_{\tilde{Y}}}{\sqrt{1+\tilde{Y}}}\psi^\mu\psi_{\mu\nu}\psi^\nu\square\psi
	\right) \; .
	\end{equation*}
	Note that the limit $D\to \infty$ translates to $\tilde{Y}\to -1$ in terms of the variable $\psi$.
		By inspecting the equations of motion following from the above action, one can notice that the first and the third terms give subdominant contributions with respect to the contribution from the second term, in the limit $\tilde{Y}\to -1$. As a result, upon redefining the function $f$, we obtain the action,
	\begin{equation}
	\label{action_NCS}
	S_{NCS}[\tilde{g}] = \frac{M_P^2}{2}\int\td^4 x \sqrt{-\tilde{g}}f\left[\psi_{\mu\nu}\psi^{\mu\nu}-(\square\psi)^2\right]  .
	\end{equation}	
	One can check that the solution for the NCS metric~(\ref{metric_infty}) and the scalar field~$\psi$, given by Eq.~(\ref{phi}) with the redefinition $\phi=q_0\psi/\sqrt{D}$, is indeed a solution for the theory~(\ref{action_NCS}), which belongs to the DHOST classes IIIa and Ia \cite{Achour:2016rkg}.
	
	\subsection{Theory giving rise to the quasi-Weyl metric}
	In this section, we discuss the theory that admits the quasi-Weyl metric of Sec.~\ref{case_-1}, which was obtained in the limit $D\to -1$, as a solution. This limit is singular for our variables, since after the time rescaling leading to the coordinates used in the metric~(\ref{metric}), the scalar field~(\ref{phi}) reads
	\begin{equation*}
\phi = \frac{q_0}{\sqrt{1+D}}\left[t + (1+D)\int \frac{\sqrt{2\tm
		r(a^2+r^2)}}{\Delta}\td r\right]\; .
	\end{equation*}
	We absorb this divergence using the same redefinition of the scalar field as in the previous section, $\phi = q_0 \psi/\sqrt{1+D}$. After this redefinition, we take the limit $D\to-1$, which results in the simple expression{\footnote{Solutions where the scalar depends only on time were discussed in Ref.~\cite{Charmousis:2015txa}.}} $\psi=t$. Taking the same limit in the general action given by the functions~(\ref{functions_general}), and expressing everything in terms of the field $\psi$, results in
\begin{equation}
\label{action_QW}
S_{QW}[\tilde{g}] = \frac{M_P^2}{2}\int\td^4 x \sqrt{-\tilde{g}}\left( \sqrt{-\tilde{Y}}f(\tilde{Y})\tilde{R} -\frac{f}{\sqrt{-\tilde{Y}}}\left[\psi_{\mu\nu}\psi^{\mu\nu}-(\square\psi)^2\right] +\frac{4f_{\tilde{Y}}}{\sqrt{-\tilde{Y}}}\psi^\mu\psi_{\mu\nu}\psi^\nu\square\psi
\right) \; ,
\end{equation}
where $f$ has also been redefined as $f\to f\sqrt{1+D}$ to absorb a residual infinite factor. One can check that the above action admits the QW metric~(\ref{metric_QW}) and $\psi=t$ as a solution. If we consider a constant $\tilde f$, the theory belongs to the Horndeski class.

	\section{Secular shifts for the EKD metric}
	\label{app:secular}
	In this appendix, we provide the expressions for secular perturbations of orbital parameters up to 2PN order in the case of the enhanced Kerr disformation, $D+1\sim \mathcal{O}(\varepsilon)$; see Sec.~\ref{case_1+D_small}. Since the EKD metric~(\ref{metric_harm}) does not fall into the class of generic $D$, we cannot use the results of Sec.~\ref{case_generic} to find the variation of orbital parameters. 
		Here we present the result of applying the osculating orbit method described in Sec.~\ref{sec:method} to the metric~(\ref{metric_harm}). We obtain
	\begin{align}
	\frac{\td \bp}{\td u}&=\frac{\la \tm^2\sin^2 \bi}{\bp(1-\ba^2 - \bb^2)^2}\left[5\ba\bb\left(5 + 2\ba^2+2\bb^2\right) + \mathcal{I}_1(\ba,\bb)\right]\nonumber\\
	&\qquad-\frac{\la^2\tm^2\ba\bb\sin^2\bi}{8\bp(1-\ba^2-\bb^2)^3}\left[-72 + \left(\ba^2+\bb^2\right)^2+6\ba^2 -2\bb^2+\left(84+(\ba^2+\bb^2)^2-6\bb^2 - 14\ba^2\right)\cos^2\bi \right]\; ,\nonumber\\
	\frac{\td \bi}{\td u}&=\frac{\la \tm^2\sin 2\bi}{4\bp^2(1-\ba^2-\bb^2)^2}\left[5\ba\bb\left(5+2\ba^2+2\bb^2\right)+\mathcal{I}_1(\ba,\bb)\right]\nonumber\\
	&\qquad +\frac{\la^2 \tm^2 \ba\bb\sin 2\bi}{32\bp^2(1-\ba^2-\bb^2)^3}\left[176 + 3\left(\ba^2+\bb^2\right)^2+86\bb^2 +78\ba^2-\left(84+(\ba^2 + \bb^2)^2-6\bb^2 -14\ba^2\right)\cos^2\bi\right]\; ,\nonumber\\
	\frac{\td \bO}{\td u} &=-\frac{3\tm\la\cos\bi}{2\bp(1-\ba^2-\bb^2)}+2\chi\left(\frac{\tm}{\bp}\right)^{3/2}+\frac{\la \tm^2 \cos\bi}{4\bp^2(1-\ba^2-\bb^2)^2}\left[48+37\ba^4 +57\bb^4 -35\ba^2 +15\bb^2+94\ba^2\bb^2 + \mathcal{I}_2(\ba,\bb)\right]\nonumber\\
	&\qquad+\frac{\la^2\tm^2\cos\bi}{32\bp^2(1-\ba^2-\bb^2)^3}\left[174 +\ba^4\left(6\bb^2-79\right)+6\ba^2\left(
	13-\bb^2 + 2\bb^4\right)+\bb^2\left(570 + 89\bb^2 + 6\bb^4\right)\right]\nonumber\\
	&\qquad+\frac{\la^2\tm^2\cos^3\bi}{32\bp^2(1-\ba^2-\bb^2)^3}\left[-66 + 2\ba^6 - 554 \bb^2 + 207\bb^4 + \ba^4\left(191+4\bb^2\right)+2\ba^2\left(\bb^4 + 207\bb^2 - 123\right)\right]\; ,\nonumber\\
	\frac{\td\ba}{\td u}&=-\frac{3\tm \bb}{4\bp(1-\ba^2-\bb^2)}\left[4\left(1-\ba^2-\bb^2\right)+\la\left(5\cos^2\bi-1\right)\right]+6\bb \chi\cos \bi\left(\frac{\tm}{\bp}\right)^{3/2}+\frac{3\tm^2\bb(10-\ba^2-\bb^2)}{4 \bp^2} 	\nonumber \\
	&\qquad + \frac{\la\tm^2}{16\bp^2(1-\ba^2-\bb^2)^2}\left[2\bb\left(-53+199\ba^2 + 34\ba^4 -19\bb^2(5+2\ba^2)-72\bb^4\right)+\sin^2\bi\,\mathcal{I}_3(\ba,\bb)+\mathcal{J}_1(\ba,\bb,\bi)\right]\nonumber\\
	&\qquad + \frac{\la\tm^2\bb\cos^2\bi}{8\bp^2(1-\ba^2-\bb^2)^2}\left[253+48\ba^4 +194\bb^4- 27\bb^4 +\ba^2(242\bb^2 -421)\right]\nonumber\\
	&\qquad -\frac{\la^2 \tm^2\bb}{64\bp^2(1-\ba^2-\bb^2)^3}\left[585+2\ba^2\left(2\ba^4 - 18\ba^2 - 147\right) +2\bb^2\left(693+34\ba^2 + 4\ba^4\right)+ 4\bb^4\left(46+\ba^2\right)\right]\nonumber\\
	&\qquad -\frac{\la^2 \tm^2\bb\cos^2\bi}{32\bp^2(1-\ba^2-\bb^2)^3}\left[-729 + \ba^2\left(290+295\ba^2+6\ba^4\right)+2\bb^2\left(6\ba^4 + 315\ba^2-733\right)+\bb^4\left(239+6\ba^2\right)\right]\nonumber\\
	&\qquad +\frac{\la^2 \tm^2\bb\cos^4\bi}{64\bp^2(1-\ba^2-\bb^2)^3}\left[-585+2\ba^2\left(-209+321\ba^2 + 4\ba^4\right)+2\bb^2\left(8\ba^4 + 936\ba^2-941\right)+2\bb^4\left(559+4\ba^2\right)\right]\; ,\nonumber\\
	\frac{\td\bb}{\td u}&=\frac{3\tm \ba}{4\bp(1-\ba^2-\bb^2)}\left[4\left(1-\ba^2-\bb^2\right)+\la\left(5\cos^2\bi-1\right)\right]-6\ba \chi\cos \bi\left(\frac{\tm}{\bp}\right)^{3/2}-\frac{3\tm^2\ba(10-\ba^2-\bb^2)}{4 \bp^2}\nonumber\\
	&\qquad +\frac{\la \tm^2}{16\bp^2(1-\ba^2-\bb^2)^2}\left[2\ba\left(87 + 46\ba^4 + 461\bb^2+ \ba^2(167+198\bb^2)+152\bb^4\right)- \sin^2\bi\, \mathcal{I}_4(\ba,\bb)-\mathcal{J}_2(\ba,\bb,\bi)\right]\nonumber\\
	&\qquad -\frac{\la \tm^2\ba\cos^2\bi}{8\bp^2(1-\ba^2-\bb^2)^2}\left[287+128\ba^4+339\bb^2+274\bb^4+\ba^2\left(-55+402\bb^2\right)\right]\nonumber\\
	&\qquad -\frac{\la^2\tm^2\ba}{64\bp^2(1-\ba^2-\bb^2)^3}\left[-429+102\ba^2 + 68\ba^4 + 2\bb^2\left(-773-18\ba^2+2\ba^4\right)+ 8\bb^4\left(\ba^2-23\right)+4\bb^6\right]\nonumber\\
	&\qquad +\frac{\la^2\tm^2\ba\cos^2\bi}{32\bp^2(1-\ba^2-\bb^2)^3}\left[-813 + 298\ba^2 + 371\ba^4 + 6\ba^6 + 2\bb^2\left(-713+375\ba^2+6\ba^4\right)+\bb^4\left(283+6\ba^2\right)\right]\nonumber\\
	&\qquad -\frac{\la^2\tm^2\ba\cos^4\bi}{64\bp^2(1-\ba^2-\bb^2)^3}\left[-909 -210\ba^2 + 762\ba^4 + 4\ba^6 + 2\bb^2\left(-821+2\ba^2(520+\ba^2)\right)+2\bb^4\left(603-2\ba^2\right)-4\bb^6\right]\label{sec_pert}\; ,
	\end{align}
	 where we defined the following functions to make the previous expressions lighter (and we use the average $\langle\cdot\rangle$ as defined in Eq.~(\ref{average})):
	\begin{equation*}
	\begin{split}
	H^{(\kappa)}(u,\ba,\bb)&= 4\sqrt{2}\left(1-\ba^2-\bb^2\right)\left(1+\ba\cos u+\bb \sin u\right)^{\kappa}\; ,\\
	\mathcal{I}_1(\ba,\bb)&=\langle \sin 2u\left(\bb \cos u - \ba \sin u\right) H^{(3/2)}(u,\ba,\bb)\rangle\; ,\\
	\mathcal{I}_2(\ba,\bb)&=4\langle \sin^2 u\left(\bb \cos u - \ba \sin u\right) H^{(3/2)}(u,\ba,\bb)\rangle\; ,\\
	\mathcal{I}_3(\ba,\bb)&=16\langle \cos u\sin^2 u\,  H^{(7/2)}(u,\ba,\bb)\rangle\; ,\\
	\mathcal{I}_4(\ba,\bb)&=16\langle \cos^2 u\sin u\,  H^{(7/2)}(u,\ba,\bb)\rangle\; ,\\
	\mathcal{J}_1(\ba,\bb,\bi)&=4\langle \left(\bb \cos u-\ba\sin u\right)H^{(3/2)}(u,\ba,\bb)\left(4\bb\cos^2\bi\sin^2 u+\sin^2\bi\sin 2u(3\ba + 4\cos u + \ba \cos 2u+\bb \sin 2u)\right)\rangle\; ,\\
	\mathcal{J}_2(\ba,\bb,\bi)&=4\langle \left(\bb \cos u-\ba\sin u\right)H^{(3/2)}(u,\ba,\bb)\left(4\ba \cos^2\bi\sin^2 u + \sin^2\bi \sin 2u\left(-3\bb -4\sin u+ \bb\cos 2u-\ba \sin 2u\right)\right)\rangle\; .
	\end{split}
	\end{equation*}
	
	Using the expressions for the secular variations of $\{\ba,\bb\}$ in Eq.~(\ref{sec_pert}), one can obtain the secular variations of $\{\bo,\be\}$. As we discussed in Sec~\ref{case_1+D_small}, see Eq.~(\ref{leading_pericenter}), the leading correction for the variation of the pericenter angle is modified due to disformality.
	Similarly to the parameters $\{\bp,\bi\}$, $\be$ receives secular corrections at 2PN order, which is one $\varepsilon$ order lower than the GR value:
	\begin{equation}
	\begin{split}
	\label{e_sec}
	\frac{\td \be}{\td u}&=\frac{\la \tm^2\sin^2 \bi}{8\bp ^2(1-\be^2)^2}\left[\be\sin 2\bo\left(17+183 \be^2+40\be^4\right) + 16\sqrt{2}\left(1-\be^2\right)^2\langle\sin2u \sin(u-\bo)\left(1+\be\cos(u-\bo)\right)^{3/2}\rangle\right]\\
	&\qquad+\frac{\be \la^2\tm^2\sin^2\bi}{64\bp^2(1-\be^2)^2}\left[4 \be^2 \sin^2 \bi \sin4\bo + 2 \sin2\bo\left(-39 + 5 \be^2 + \be^4 + \cos^2\bi\left(-81-25\be^2 + \be^4\right)\right)\right]\; .
	\end{split}
	\end{equation}
	Though the secular contributions for $\{\bp,\bi,\be\}$ appear at a lower PN order than for the Kerr metric, one can show that over long timescales these contributions average out to 0. Indeed, the characteristic timescale over which the pericenter angle $\bo$ varies is shorter than for other parameters, as the leading secular shift for $\bo$ appears at 1PN order. Hence, one can average the secular variations of orbital parameters over $\bo$, while keeping other parameters fixed. It is not necessary to perform another 2-timescale analysis since the terms containing $\bo$ are already of 2PN order, which means such an analysis would only be relevant if we were interested in higher-order PN terms (see Ref.~\cite{Will:2016pgm}). From Eq.~(\ref{e_sec}), one can easily see that
	\begin{equation}
	\frac{1}{2\pi}\int_{0}^{2\pi}\frac{\td\be}{\td u}\td\bo=0
	\end{equation}
	at this PN order, and hence the variation of eccentricity averages out to 0 over a long timescale. One can check that the same is true for the parameters $\{\bp,\bi\}$.
	%%%%%%%%%%%%%%%%%%%%%%%%%%%%%%%%%%%%%%%%%%%%%%%%%%%
	\section{Leading-order pericenter precession using the textbook method}
	\label{app:textbook}
	In this appendix, we apply the textbook method to derive the leading term for the pericenter precession for the EKD metric considered in Sec.~\ref{case_1+D_small}, and compare it with the result obtained in Eq.~(\ref{leading_pericenter}).
		Assuming that the trajectory is in the equatorial plane $\theta=\pi/2$, one can write first-order geodesic equations.
		This is not the case for the general motion outside of the equatorial plane, because of the absence of a nontrivial Killing tensor for the disformed Kerr metric.
	The orbit of stars around the central black hole of the Galaxy is approximately an ellipse, for which the energy and angular momentum can be written as
	\begin{align*}
	E^2 &\simeq 1 - \frac{\tilde{M}}{A}\; ,\\
	L^2&\simeq A \tilde{M}(1-e^2)\; ,
	\end{align*}
	where $A$ and $e$ are respectively the semimajor axis and eccentricity of the Kepler orbit.
	We now combine the geodesic equations for the variables $\{r,\vp\}$, and use the standard variable $U\equiv 1/r$. 
	By substituting the above values for energy and angular momentum in the geodesic equations, a second-order equation follows:
	\begin{equation}
	\label{usecond}
	U''(\vp) + F(U,\tilde{M}, a,D)=0\; ,
	\end{equation} 
	where $F$ is a complicated expression which also depends on $A$, $e$.  Following the standard procedure (see e.g. Ref.~\cite{dInverno:1992gxs}), we introduce a small parameter
	\begin{equation}
	\eta = \frac{3\tilde{M}^2}{L^2} \simeq \frac{3}{(1-e^2)}\varepsilon\; ,
	\end{equation}
	where $\varepsilon=\tm/A$ is the small parameter used throughout the main text.
	We again assume $\tilde{a} = \chi \tilde{M}$ with $\chi\sim\mathcal{O}(1)$, so that we can also express $a$ in terms of $\eta$. 
	We look for a solution to Eq.~(\ref{usecond}) of the form
	\begin{equation*}
	U = \frac{1+ e\cos\vp}{A(1-e^2)}+\eta\; \delta U(\vp)\; ,
	\end{equation*}
	where the first part corresponds to the Kepler orbit, and the second term is the leading correction in $\eta$.
	For the Schwarzschild metric ($a=0$), this leads to the equation
	\begin{equation*}
	\delta U_S'' + \delta U_S = \frac{(1+e\cos\vp)^2}{A(1-e^2)}\; .
	\end{equation*}
	The above equation can be solved (see e.g. Ref.~\cite{dInverno:1992gxs}) to write down the solution for $U$ at first order in $\eta$:
	\begin{equation*}
	U_S \simeq \frac{1+e\cos\left[\vp(1-\eta)\right]}{A(1-e^2)}\; .
	\end{equation*}
	From this expression, one can calculate the precession of the pericenter $\Delta\Phi_S$ as
	\begin{equation}
	\label{precession_S}
	\Delta\Phi_S = 2\pi\left(\frac{1}{1-\eta}-1\right)\simeq 2\pi\eta=\frac{6\pi\tilde{M}}{A(1-e^2)}
	\end{equation}
	The asymptotic expression of the disformed metric suggests that for the EKD metric, i.e. when $D+1\sim \mathcal{O}(\varepsilon)$, corrections from higher-order terms in $\varepsilon$ become of 1PN order, i.e. comparable to the leading Schwarzschild corrections. To check this, we rewrite Eq.~(\ref{assumptions_Dneg}) in terms of $\eta$:
	\begin{equation*}
	D=-1 + \frac{(1-e^2)\chi^2}{3\la}\eta\; .
	\end{equation*}
	Assuming this form for $D$ in Eq.~(\ref{usecond}) and expanding to $\mathcal{O}(\eta)$, we obtain the following equation for the correction to the Kepler orbit:
	\begin{equation*}
	\delta U'' + \delta U = \frac{(1+e\cos\vp)^2}{A(1-e^2)} + \frac{\la(1+e\cos\vp)(2 + e\cos\vp + e^2 - 2e^2\cos(2\vp))}{3A(1-e^2)^2}\; .
	\end{equation*}
	Solving this equation and keeping only the terms of the form $\vp\sin\vp$ which provide a secular shift, the standard procedure gives,
	\begin{equation}
	\label{precession_textbook}
	\Delta\Phi =\Delta\Phi_S\left(1+\frac{\la}{2 (1-e^2)}\right)\; .
	\end{equation}
	Thus for a spinning EKD black hole the corrections to the pericenter precession due to modifications of gravity are of the same order as the leading GR effect. 
		Equation~(\ref{precession_textbook}) provides a useful check for our calculations in the main text, where we obtained secular shifts for the EKD black hole by the orbital perturbation method. In particular, we obtained the expression for the pericenter precession in Eq.~(\ref{leading_pericenter}). The two expressions agree after setting $\bi=0$ in Eq.~(\ref{leading_pericenter}), which corresponds to an orbit in the equatorial plane, as assumed in this section.
	
	\section{Comparison to other metrics}
	\label{app_others}
	It is instructive to compare the Kerr disformation to other metrics presented in the literature, in the asymptotic regime, i.e. for large distances.
		The main challenge here is that normally line elements are written in different coordinates, which makes a direct comparison impossible.
		We will write all the interesting metrics in Boyer-Lindquist-like coordinates, so that we can see a connection to our asymptotic expansion of the disformed Kerr metric~(\ref{metric_as}).
	
	We first explain how the expression for the Butterworth-Ipser metric in Boyer-Lindquist-like coordinates can be obtained in Eq.~(\ref{BI_BL}). 
		We start with the ansatz of Ref.~\cite{BW71} for a circular and axisymmetric metric in  quasi-isotropic coordinates:
	\begin{equation*}
	\td s^2 = -e^{2\nu}\td t^2 + e^{2\psi}\left(\td \vp -\tilde{\omega} \td t\right)^2 +e^{2\mu}\left(\td R^2 + R^2 \td\theta^2\right)\; ,
	\end{equation*} 
	where $\nu$, $\psi$, $\tilde{\omega}$ and $\mu$ are functions of $R$ and $\theta$. The BI metric at 2PN order is given by the following expressions (see Ref.~\cite{Friedman:2013xza} for instance\footnote{There is a typo in Eq.~(3.29c) of this reference, where the quantity $\mu_S$ should be added to the right-hand side. We use the notation $a_0$ instead of $a$ to avoid confusion with the Kerr spin parameter.}):
	\begin{equation}
	\begin{split}
	\nu &= -\frac{\tm}{R}+\left[-\frac{1}{12}+a_0-\left(4a_0+q\right)P_2(\cos\theta)\right]\frac{\tm^3}{R^3}+\mathcal{O}(R^{-4})\; ,\\
	\mu &= \frac{\tm}{R} - \left[\frac{1}{4}+a_0 -4a_0 P_2(\cos\theta)\right]\frac{\tm^2}{R^2} + \mathcal{O}(R^{-3})\; ,\\
	\psi &= \log(R\sin\theta) +\frac{\tm}{R}+ \left[3 a_0 - \frac{1}{4}\right]\frac{\tm^2}{R^2}+ \mathcal{O}(R^{-3})\; ,\\
	\tilde{\omega} &= \frac{2\chi\tm^2}{R^3} + + \mathcal{O}(R^{-4})\; ,
	\end{split}
	\end{equation}
	where $P_2(x)= (3x^2-1)/2$ is a Legendre polynomial.
	The coordinate transformation that brings the disformed Kerr metric to quasi-isotropic coordinates $\{t,R,\theta,\vp\}$ at 2PN order reads:
	\begin{equation}
	\label{QItoBL}
	r = R\left[1+\frac{\tm}{2R}\left(1+\frac{\chi}{\sqrt{1+D}}\right)\right]\left[1+\frac{\tm}{2R}\left(1-\frac{\chi}{\sqrt{1+D}}\right)\right]\; .
	\end{equation}	
	In the case $D=0$, the previous redefinition brings the full Kerr metric to quasi-isotropic coordinates: see e.g. Ref.~\cite{Lanza_1992}.
For nonzero disformal parameter $D$, one must introduce extra factors containing $D$ in the transformation, and the resulting metric is only quasi-isotropic asymptotically. By inverting the above relation, we obtain the expression~(\ref{BI_BL}) for the BI metric in BL-like coordinates.

	A well-known example  of an axisymmetric and stationary spacetime is the Hartle-Thorne (HT) metric, that describes a metric for slowly rotating stars.
		Note that the ``quasi-Kerr'' metric \cite{Glampedakis:2005cf}, sometimes used as an alternative to the Kerr metric,  is exactly the HT metric up to 2PN order.
	The HT metric for slowly rotating stars can be written in a form which reduces to BL coordinates in the Kerr limit. At 2PN order, this metric reads
	\begin{equation}
	\label{HT2PN}
	\begin{split}
	\td s^2_\text{HT}&=-\left(1-\frac{2\tm}{r}-\frac{\tm^3}{r^3}\left[q_\text{HT}-\chi^2 + (\chi^2 - 3q_\text{HT})\cos^2\theta\right]\right)\td t^2 + \left(1+\frac{2\tm}{r}+\frac{\tm^2(4-\chi^2\sin^2\theta)}{r^2}\right)\td r^2\\
	&+r^2\left(1+\frac{\tm^2\chi^2\cos^2\theta}{r^2}\right)\td\theta^2+r^2\sin^2\theta\left(1+\frac{\tm^2\chi^2}{r^2}\right)\td\vp^2 -\frac{4\tm^2\chi\sin^2\theta}{r}\td t\td\vp\; .
	\end{split}
	\end{equation}
	We have used the expression of the metric in the appendix of Ref.~\cite{Hartle:1968si}\footnote{Notice a typo in the coordinate transformation to BL-like coordinates (as pointed out in Ref.~\cite{Glampedakis:2005cf}).} with the identification $Q=q_\text{HT}\tm^3$ and $J=\chi \tm^2$.
	The HT metric up to 2PN order is a subclass of the BI metric. This can be seen by comparing Eqs.~(\ref{HT2PN}) and~(\ref{BI_BL}) (where one sets $D=0$) with the identification $q=-q_\text{HT}$ and $a_0= \chi^2/12$.
The disformed metric given by Eq.~(\ref{metric_as}) cannot be matched with Eq.~(\ref{HT2PN}), as can be seen from a direct comparison. 
		The other variants of the Kerr disformation cannot be matched to the HT metric either. 
		In particular, for NCS, EKD and \emph{e}KD this follows from the fact that these cannot be matched to the BI metric. 
		The case of the QW metric can be checked by comparing the different elements of the metric, taking into account that the $(t\varphi)$ component in the QW metric is absent.
		On the other hand, the orbital shifts up to 2PN order in the generic case~(\ref{sec_general}) can be matched to the orbital shifts for the HT metric, by identifying $q_\text{HT}=\chi^2/(1+D)$. Similarly, the secular variations for the NCS metric (the limit of large $D$) at 2PN are identical to those for HT metric when setting $q_\text{HT}=0$.
		In order to recover the 2PN secular shifts of the QW metric, Eq.~(\ref{sec_formal}), one sets $q_\text{HT}=\chi_1^2$ and $\chi=0$ in the metric~(\ref{HT2PN}). 
	
	Another spacetime worth comparing our results to is the Johannsen metric \cite{Johannsen:2015pca}. Unlike the Hartle-Thorne metric, it describes a rotating black hole. Using the notations of Ref.~\cite{Johannsen:2015pca}, the line element at 2PN order in BL coordinates reads
	\begin{equation}
	\label{J2PN}
	\begin{split}
	\td s^2_\text{J}&=-\left(1-\frac{2\tm}{r}+\frac{\tm^3}{r^3}\left[2\alpha_{13}-\epsilon_3-2\chi^2\cos^2\theta\right]\right)\td t^2 + \left(1+\frac{2\tm}{r}+\frac{\tm^2(4-\alpha_{52}-\chi^2\sin^2\theta)}{r^2}\right)\td r^2\\
	&+r^2\left(1+\frac{\tm^2\chi^2\cos^2\theta}{r^2}\right)\td\theta^2+r^2\sin^2\theta\left(1+\frac{\tm^2\chi^2}{r^2}\right)\td\vp^2 -\frac{4\tm^2\chi\sin^2\theta}{r}\td t\td\vp\; .
	\end{split}
	\end{equation}
	Note that the Johannsen metric is circular by construction. 
	The Kerr metric at this order is obtained by setting $\alpha_{13}=\alpha_{52}=\epsilon_3=0$ in the above expression. 
	In general, the metric~(\ref{J2PN}) does not match the BI metric at 2PN order.  
		Only for a special combination of the parameters, $q=-\chi^2$, $a_0=\chi^2/12$, $\alpha_{52}=0$ and $\epsilon_3 = 2\alpha_{13}$, do these two metrics match. 
		In this case the coordinate-invariant quadrupole is the same as for Kerr and the no-hair theorem is not violated at this order. 
		The Johannsen metric can also be mapped to other Kerr-like metrics; see Ref.~\cite{Johannsen:2015pca} for details.
		One can also verify by comparing Eqs.~(\ref{J2PN}) and~(\ref{metric_as}) that the disformed metric in the generic case does not match the Johannsen metric at this order. 
		The same conclusion is true for the other variants of the disformed metric.
		As for the secular variation of orbital parameters, the 2PN Schwarzschild terms are modified for the Johannsen metric with respect to the Kerr case, while the quadrupole terms proportional to $\chi^2$ are the same as in the Kerr case (Eq.~(\ref{sec_general}) with $D=0$). Notice that this is different from the generic case of the disformal parameter, where we found that the terms proportional to $\chi^2$ are modified with respect to the Kerr case.
		
%\bibliographystyle{JHEP}
%\bibliography{biblio}

\providecommand{\href}[2]{#2}\begingroup\raggedright\endgroup

\end{document}